\documentclass[12pt]{article}
\usepackage{verbatim}
\usepackage[dvipsnames]{xcolor} 
\usepackage{graphics,amsmath,pstricks}
\usepackage{amssymb,enumerate}
\usepackage{amsbsy,amsmath,amsthm,amsfonts, amssymb}
\usepackage{graphicx, rotate, array}
\usepackage{multirow}
\usepackage{float}
\usepackage{natbib}

\usepackage{mathrsfs}
\usepackage{algorithm} 
\usepackage{algpseudocode} 
\usepackage[english]{isodate} 
\usepackage{multirow}
\cleanlookdateon
\newtheorem{lemma}{Lemma}
\newtheorem{theorem}{Theorem}
\usepackage{listings}
\usepackage{subcaption} 
\usepackage{booktabs} 
\usepackage{bbm}
\usepackage{xr-hyper,xr}
\usepackage{hyperref}
\usepackage{ifthen}
\usepackage{pstricks}
\usepackage{scalerel}[2016/12/29]
\newcommand\lilsigma{\scaleobj{0.75}{\Sigma}}
\cleanlookdateon

\newtheorem{defi}{Definition}[theorem]
\newtheorem{lem}{Lemma}
\newtheorem{thm}{Theorem}

\graphicspath{ {images/} }

\newboolean{DEBUG}
\setboolean{DEBUG}{true}

\newrgbcolor{amycolor}{.5 .1 .99} 
\ifthenelse {\boolean{DEBUG}}
{\newcommand{\amy}[1]{{\amycolor{[\@Amy: #1]}}}}
{\newcommand{\amy}[1]{}}

\newrgbcolor{davidcolor}{.1 .7 .1}
\ifthenelse {\boolean{DEBUG}}
{\newcommand{\david}[1]{{\davidcolor{[\@David: #1]}}}}
{\newcommand{\david}[1]{}}

\begin{document}

\def\spacingset#1{\renewcommand{\baselinestretch}%
{#1}\small\normalsize} \spacingset{1.1}

\title{\bf Modeling complex measurement error in microbiome experiments to estimate relative abundances and detection effects}
  \author{David S Clausen and
    Amy D Willis\thanks{
      The authors gratefully acknowledge support from the National Institute of General Medical Sciences (R35 GM133420). Correspondence: adwillis@uw.edu} \\
    Department of Biostatistics, University of Washington}
  \maketitle

\begin{abstract}
Accurate estimates of microbial species abundances are needed to advance our understanding of the role that microbiomes play in human and environmental health. However, artificially constructed microbiomes demonstrate that intuitive estimators of microbial relative abundances are biased. To address this, we propose a semiparametric method to estimate relative abundances, species detection effects, and/or cross-sample contamination in microbiome experiments. We show that certain experimental designs result in identifiable model parameters, and we present consistent estimators and asymptotically valid inference procedures. Notably, our procedure can estimate relative abundances on the boundary of the simplex. We demonstrate the utility of the method for comparing experimental protocols, removing cross-sample contamination, and estimating species' detectability.
\end{abstract}

\section{Introduction}

\label{sec:intro}

Next generation sequencing (NGS) has profoundly impacted the study of microbial communities (microbiomes), which reside in the host-associated environments (such as the human body) as well as natural environments (such as soils, salt- and freshwater systems, and aquifers). 
NGS identifies contiguous genetic sequences, which can be grouped and counted to provide a measure of their abundances. 
However, even when sequences can be contextualized using databases of microbial DNA, NGS measurements of microbial abundances are not proportional to the abundances of organisms in their originating community. In this paper, we consider the problem of recovering microbial relative abundances from noisy and distorted NGS abundance measurements.  

Two main avenues of statistical research have considered the analysis of microbial abundance data from NGS:
batch effects removal and differential abundance. 
Batch effects are systemic distortions in observed abundance data due either to true biological variation (e.g., cage/tank effects) or measurement error (e.g., variation in extraction kits). Tools to remove batch effects from microbiome data include both methods adapted from RNA-seq and microarray analysis (e.g., \citep{leek2007capturing, johnson2007adjusting, gagnon2012using, sims2008removal}) as well as microbiome-specific approaches \citep{gibbons2018correcting, dai2019batch}. 
A key motivation for batch effect removal is to allow for more precise comparisons of biological differences between microbial communities, and methods for ``differential abundance'' offer one avenue for assessing difference. Differential abundance methods typically aim to detect microbial units (such as species or gene abundances) that are present in the data at different average levels across groups. Differential abundance methods differ widely in generative models for the data \citep{love2014moderated,Martin:2019wj,li2021ifaa} and/or their approaches to transforming the data before performing regression analyses \citep{fernandes2014unifying,mandal2015analysis,mallick2021multivariable}.

In this paper, we consider a distinct goal from batch effects removal and differential abundance: modeling the relationship between NGS measurements and the true microbial composition of the community that was sequenced. Our methodology can estimate relative abundances of individual microbial species in specimens by correcting for the unequal detectability of microbial species, accounting for unequal depths of sequencing across samples, and removing batch-specific contamination in samples. 
In addition, it can estimate species detectabilities, sample intensities, and cross-sample contamination. 
Our model incorporates information about shared origins of samples (e.g., replicates and dilution series), shared processing of samples (e.g., sequencing or DNA extraction batches) and known information about sample composition (e.g., mock communities and reference protocols), allowing us to identify and estimate statistical parameters that are not accounted for in most models for microbial abundances (e.g., species detectabilities and contamination). While methods exist to estimate either species detectabilities \citep{McLaren:2019cn, silverman2021measuring, zhao2021log} or contamination  \citep{knights2011bayesian,shenhav2019feast,davis2018simple}, to our knowledge, no methods can simultaneously estimate species detectabilities, true microbial relative abundances, sample intensities, and contamination. 

Because we consider estimation of relative abundances, and we do \textit{not} assume that all species are present in all samples, our method entails estimation of a parameter that may fall on the boundary of its parameter space. 
We use estimating equations to construct our estimator, prove semiparametric consistency and existence of limiting laws, construct a highly stable algorithm to find a solution that may fall on the boundary, and develop uncertainty quantification that accounts for the boundary problem. Therefore, in addition to its scientific contribution, our methodology employs contemporary statistical, probabilistic, and algorithmic tools.

We begin by constructing our mean model (Section 2), and discussing identifiability (Section 3), estimation (Section 4), and asymptotic guarantees (Section 5). We demonstrate applications of our method to relative abundance estimation, contamination removal, and the comparison of multiple experimental protocols (Section 6), and evaluate its performance under simulation (Section 7). We conclude with a discussion of our approach and areas for future research (Section 8). Software implementing the methodology are implemented in a \texttt{R} package available at \url{https://github.com/statdivlab/tinyvamp}. Code to reproduce all simulations and data analyses are available at \url{https://github.com/statdivlab/tinyvamp_supplementary}.

\section{A measurement error model for microbiome data}


Here we propose a model to connect observed microbial abundances to true relative abundances. 
We use the
term ``read'' to refer to the observed measurements (which is \textit{not} required to be integer-valued),  ``taxon'' to the categorical unit under study (e.g., species, strains or cell types), 
``sample'' for the unit of sequencing, and ``specimen'' for the unique source of
genetic material (which may be repeatedly sampled). 



\begin{figure}
\begin{centering}
\includegraphics[width=\textwidth]{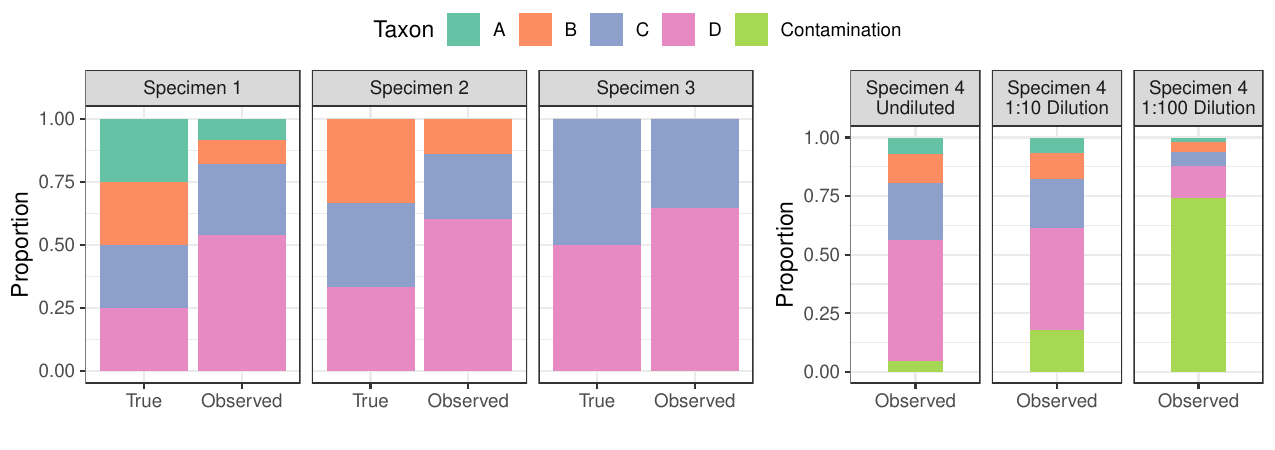}
 \caption{Two features of microbiome measurement data that are accounted for in our model include (left) observed relative abundances are predictably biased relative for true relative abundances; and (right) contamination is associated with biomass. Here the proportion of reads due to contamination is larger in lower-biomass than in higher-biomass samples, reflecting the greater impact of an approximately constant amount of contamination on lower biomass samples. 
 } \label{fig:motivation}
\end{centering}
\end{figure}

Let $W_i = (W_{i1},\dots, W_{iJ})$ denote observed reads from taxa $1, \dots, J$ in sample $i$. A common modeling assumption for high-throughput sequencing data is
\begin{align}
\mathbb{E}[W_i|\mathbf{p}_i, \gamma_i] &= \text{exp}(\gamma_i)\mathbf{p}_i
\end{align}
where $\mathbf{p}_i \in \mathbb{S}^{J-1}$ is the unknown relative abundances of taxa $1, \dots, J$ and $\text{exp}(\gamma_i) \in \mathbb{R}^{+}$ is a sampling intensity parameter (throughout, we let $\mathbb{S}^{J-1}$ denote the closed $J-1$-dimensional simplex). 
Unfortunately, microbial taxa are not detected equally well (Figure \ref{fig:motivation} (left); \cite{McLaren:2019cn} and references therein). To account for this, we begin by considering the model
\begin{align} \label{eff_single_sample}
\mathbb{E}[W_i|\mathbf{p}_i, \beta, \gamma_i] &=
\text{exp}(\gamma_i)(\text{exp}(\beta) \circ \mathbf{p}_i)
\end{align}
where $\beta = \beta_1, \dots, \beta_J$
represents the detection effects for taxa $1, \dots, J$ in a given experiment ($\circ$ indicates element-wise multiplication). As an identifiability constraint, we set $\beta_J =0$, and so interpret $\exp(\beta_j), j = 1, \ldots, J$,
as the degree of multiplicative over- or under-detection of taxon $j$ relative to taxon $J$. We discuss identifiability in more detail in Section \ref{sec:identifiability}.

We now generalize model (\ref{eff_single_sample}) to
multiple samples and one or more experimental protocols.
For a study involving $n$ samples of $K$ unique specimens ($n \geq K$), we define the \textit{sample design matrix} $\mathbf{Z} \in \mathbb{R}^{n \times K}$ to link samples to specimens ($\mathbf{Z}_{i \cdot} \in \mathbb{S}^{K - 1}$). In most experiments, $Z_{ik} = \mathbf{1}_{\{\text{sample $i$ taken from specimen $k$}\}}$, but more complex designs are also possible (e.g., $Z_{ik} = Z_{ik'} = \frac{1}{2}$ denotes that sample $i$ is a 1:1 mixture of specimens $k$ and $k'$). 
Letting $\mathbf{p}$ be a $K \times J$ matrix such that the $k$-th row of $\mathbf{p}$ gives the true relative abundances of taxa in specimen $k$, we have that the relative abundance vector for sample $i$ is $(\mathbf{Z}\mathbf{p})_{i \cdot}$. 
We allow differing detections across samples to be specified via the \textit{detection design matrix} $\mathbf{X} \in \mathbb{R}^{n\times p}$. For example,
if samples are processed using one of $p$ different protocols, we might specify $X_{iq} = \mathbf{1}_{\{\text{sample $i$ processed with protocol $q$}\}}$. 
Accordingly, we now consider a detection effect matrix $\boldsymbol{\beta} \in \mathbb{R}^{p \times J}$, with $\boldsymbol{\beta}_{\cdot J} = \mathbf{0}_{p}$ as before.
Therefore, one generalization of model (\ref{eff_single_sample}) is
\begin{align} \label{model:nonspurious_mean}
\mathbb{E}\left[\text{reads due to contributing samples}|\boldsymbol{\beta},
\mathbf{p},\boldsymbol{\gamma},
\mathbf{Z},\mathbf{X}\right] & = \mathbf{D}_{\boldsymbol{\gamma}}
\left(\mathbf{Z}\mathbf{p}\right) \circ \text{exp}
\left( \mathbf{X}\boldsymbol{\beta}\right),
\end{align}
where $\mathbf{D}_{\boldsymbol{\gamma}} = \text{diag}(\text{exp}(\boldsymbol{\gamma}))$, and
where exponentiation is element-wise.

We now extend this model to reflect contributions of contaminant sources  (Figure \ref{fig:motivation} (right)). We consider $\tilde{K}$ sources of contamination with relative abundance profiles given in the rows of $\tilde{\mathbf{p}} \in \mathbb{R}^{\tilde{K} \times J}$.
To link sources of contamination to samples, we let $\tilde{\mathbf{Z}} \in \mathbb{R}^{n \times \tilde{K}}$ be a \textit{spurious read design} matrix. Most commonly we expect $\tilde{Z}_{i\tilde{k}} = \mathbf{1}_{\{\text{source $\tilde{k}$ may contribute reads to sample $i$}\}}$, but we give an example of an analysis with more complex $\tilde{\mathbf{Z}}$ in Section \ref{subsec:karstens}.
Then, along with contaminant read intensities
$\tilde{\boldsymbol{\gamma}} = [\tilde{\gamma}_1,\dots, \tilde{\gamma}_{\tilde{k}}]^{\text{T}}$, we propose to model
\begin{align} \label{model:spurious_mean}
\mathbb{E}\left[\text{reads due to spurious sources}|\boldsymbol{\gamma},
\tilde{\boldsymbol{\gamma}},
\tilde{\mathbf{p}},
\tilde{\mathbf{Z}}\right] & = \mathbf{D}_{\boldsymbol{\gamma},\boldsymbol{\alpha}} \tilde{\mathbf{Z}}\left[\tilde{\mathbf{p}}\circ \text{exp}\left(\tilde{\boldsymbol{\gamma}}\mathbf{1}^{\text{T}}_J\right)\right].
\end{align}  
Here, $\mathbf{D}_{\boldsymbol{\gamma},\boldsymbol{\alpha}}$ is a diagonal matrix with diagonal elements equal to $\exp(\boldsymbol{\gamma} + \mathbf{V}\boldsymbol{\alpha})$ for 
design matrix $\mathbf{V} \in \mathbbm{R}^{n\times K_{\alpha}}$ and parameter $\boldsymbol{\alpha} \in \mathbbm{R}^{K_{\alpha}}$. As discussed in more detail in Section \ref{subsec:karstens}, $\mathbf{V}\boldsymbol{\alpha}$ allows the intensity of contamination to vary across sources, but 
typically $\boldsymbol{\alpha} = \mathbf{0}_{K_\alpha}$.  
Additionally, while we could incorporate a detection design matrix for contaminant reads (replacing (\ref{model:spurious_mean}) with
$\mathbf{D}_{\boldsymbol{\gamma},\boldsymbol{\alpha}}\tilde{\mathbf{Z}} \left[\tilde{\mathbf{p}} \circ \text{exp}\left(\tilde{\boldsymbol{\gamma}}\mathbf{1}_J^T + \tilde{\mathbf{X}}\boldsymbol{\beta}\right)\right]$ for $\tilde{\mathbf{X}} \in \mathbb{R}^{\tilde{K} \times p}$),
for most practical applications it is sufficient to identify $\tilde{\mathbf{p}}$ up to detection effects.
Therefore, combining models (\ref{model:nonspurious_mean}) and (\ref{model:spurious_mean}), we propose
the following mean model for next-generation sequencing data $\mathbf{W} \in \mathbb{R}^{n \times J}$:
\begin{align}
\boldsymbol{\mu} :=\mathbb{E}\left[\mathbf{W}|\boldsymbol{\beta}, \mathbf{p}, \boldsymbol{\gamma}, \tilde{\mathbf{p}}, \tilde{\boldsymbol{\gamma}}\right]
=\mathbf{D}_{\boldsymbol{\gamma}}\left(\mathbf{Z}\mathbf{p}\right) \circ \text{exp}\left( \mathbf{X}\boldsymbol{\beta}\right) +
\mathbf{D}_{\boldsymbol{\gamma},\boldsymbol{\alpha}}\tilde{\mathbf{Z}} \left[\tilde{\mathbf{p}} \circ \text{exp}\left(\tilde{\boldsymbol{\gamma}}\mathbf{1}_J^T 
\right)\right]. \label{mean_model}
\end{align}

\section{Model identifiability} \label{sec:identifiability}

Our highly flexible mean model \eqref{mean_model} encompasses a wide variety of experimental designs and targets of estimation. As a result, without additional knowledge, the parameters in the mean model \eqref{mean_model} will generally be unidentifiable. 
We provide a general condition for identifiability in SI Lemma 1, and prove identifiability the following specific cases:
\begin{enumerate}[(a)]
  \item Estimating relative detectability under different sampling protocols using specimens of unknown composition (see SI Section 9.2 for identifiability; Section \ref{subsec:costea} for analysis) 
  \item Estimating contamination using dilution series (see SI Section 9.3 for identifiability; Section \ref{subsec:karstens} for analysis)  
  \item Estimating contamination, detectability and composition when some samples have known composition (see SI Section 9.4 for identifiability; Section \ref{simulations} for analysis) 
\end{enumerate}
In addition to knowledge of $\mathbf{X}$, $\mathbf{Z}$ and $\tilde{\mathbf{Z}}$, constraints on the entries of $\boldsymbol{\beta}$ and/or $\mathbf{p}$ can result in the identifiability of model parameters. For example, in (c), some rows of $\mathbf{p}$ are known. As mentioned previously, an identifiability constraint on $\beta$ (e.g., $\boldsymbol{\beta}_{\cdot J} = \mathbf{0}_p$) is always necessary.

\section{Estimation and optimization}

We propose to estimate parameters
$\boldsymbol{\theta}^{\star} := (\boldsymbol{\theta}, \boldsymbol{\gamma}) :=$ $(\boldsymbol{\beta}, \mathbf{p},\tilde{\mathbf{p}}, \tilde{\boldsymbol{\gamma}},\boldsymbol{\gamma},\boldsymbol{\alpha})$ 
as M-estimators. 
We use likelihoods to define estimating equations but do \textit{not} require or assume that the
distribution of $\mathbf{W}$ lies in any particular parametric class. We show
consistency and weak convergence of our estimators of $\boldsymbol{\theta} := (\boldsymbol{\beta}, \mathbf{p}, \tilde{\mathbf{p}}, \tilde{\boldsymbol{\gamma}},\boldsymbol{\alpha})$
under
mild conditions in SI Section 10. 
We use $\boldsymbol{\theta}_0$ to denote the true value of $\boldsymbol{\theta}$.
Our unweighted objective is given by a Poisson log-likelihood:
\begin{align}
M^{\star}_n(\boldsymbol{\theta}^{\star}) := \frac{1}{n}l_n(\boldsymbol{\theta}^{\star}) = \frac{1}{n}\mathbf{1}^T\left[\text{vec}\left(\textbf{W}\right) \circ \text{log}\left(\text{vec}\left({\boldsymbol{\mu}\left(\boldsymbol{\theta}^{\star}\right)}\right)\right) - \text{vec}\left(\boldsymbol{\mu}\left(\boldsymbol{\theta}^{\star}\right)\right)\right].
 \end{align}
We use $M_n(\boldsymbol{\theta})$ to indicate the profile log-likelihood $\text{sup}_{\boldsymbol{\gamma} \in \mathbb{R}^n} M^{\star}_n(\boldsymbol{\theta}, \boldsymbol{\gamma})$, considering the elements of $\boldsymbol{\gamma} \in \mathbb{R}^n$ as sample-specific nuisance parameters. 
Consistency of $\hat{\boldsymbol{\theta}}$ for $\boldsymbol{\theta}_0$ does not require
$W_{ij}$ to follow a Poisson distribution, however, the proposed estimator
will be inefficient if the relationship between $\mathbbm{E}[W_{ij}|Z_i,X_i,\tilde{Z}_i, \gamma_i, \boldsymbol{\theta}_0]$
and $\text{Var}[W_{ij}|Z_i,X_i,\tilde{Z}_i, \gamma_i, \boldsymbol{\theta}_0]$ is not linear \citep{mccullagh1983quasi}.
Therefore, to obtain a more efficient estimator, we also
consider maximizing a reweighted Poisson log-likelihood, with weights chosen on the basis
of a flexibly estimated mean-variance relationship.
We motivate our specific choice of weights via 
the Poisson score equations
(see SI Section 9).
%
%
For weighting vector $\hat{\mathbf{v}} \in \mathbb{R}_{+}^{nJ}$ such that $\mathbf{1}^T\hat{\mathbf{v}} = nJ$,
we define the weighted Poisson log-likelihood as
\begin{align}
M_n^{\star\hat{\mathbf{v}}}(\boldsymbol{\theta}^{\star}) :=  \frac{1}{n}\hat{\mathbf{v}}^T \left[\text{vec}\left(\textbf{W}\right) \circ \text{log}\left(\text{vec}\left({\boldsymbol{\mu}\left(\boldsymbol{\theta}^{\star}\right)}\right)\right) - \text{vec}\left(\boldsymbol{\mu}\left(\boldsymbol{\theta}^{\star}\right)\right)\right].
 \end{align}
We define $M^{\hat{\mathbf{v}}_n}_n(\boldsymbol{\theta})$ by analogy with $M_n(\boldsymbol{\theta})$ above. We select $\hat{\mathbf{v}}$ via a centered isotonic regression \citep{oron2017centered} of squared residuals
$\text{vec}\big[(\mathbf{W} - \boldsymbol{\hat{\mu}})^2\big]$ on fitted means $\text{vec}\big[\boldsymbol{\hat{\mu}}\big]$
obtained from the unweighted objective. Full details are given in SI Section 10, but briefly, we set $\hat{v}_{ij} \propto \frac{\hat{\mu}_{ij} + 1}{\hat{\sigma}^2_{ij} + 1}$, where $\hat{\sigma}^2_{ij}$ is the
monotone regression fitted value for $\mu_{ij}$.

The estimators defined by optima of the weighted or unweighted Poisson likelihoods
given above are consistent for the true value of $\boldsymbol{\theta}$ and converge weakly to well-defined
limiting distributions at $\sqrt{n}$ rate, though the form of this distribution in general depends on the
true value of $\boldsymbol{\theta}$.
We leverage an approach from \citet{van2000asymptotic} to prove consistency,
and we combine a bracketing
argument with a directional delta method theorem of \citet{dumbgen1993nondifferentiable} to show
weak convergence.
Proofs are given in SI Section 11.

Computing maximum (weighted) likelihood estimates 
of
$\boldsymbol{\theta}^{\star}$ is a constrained optimization problem,
as the relative abundance parameters in our model are simplex-valued, and the estimate may lie on the boundary of the simplex. Therefore, we minimize $f_n(\boldsymbol{\theta}^{\star}) = -M^{\star}_n(\boldsymbol{\theta}^{\star})$ or $f_n(\boldsymbol{\theta}^{\star}) = -M^{\star \hat{\mathbf{v}}_n}_n(\boldsymbol{\theta}^{\star})$ in two steps.
%
%
In the first step, we employ the barrier method,
converting our constrained optimization
problem into a sequence of unconstrained optimizations, permitting solutions
progressively closer to the boundary. That is, for
barrier penalty parameter $t$, we update  $\boldsymbol{\theta}$ as
\begin{align} \label{orig_par_penalty}
\boldsymbol{\theta}^{\star(r + 1)} &= \text{arg min}_{\boldsymbol{\theta}} \left(f_n(\boldsymbol{\theta}^{\star}) +
\frac{1}{t^{(r)}} \left[
\sum_{k=1}^K \sum_{j=1}^J -\text{log} p_{kj}  +
\sum_{\tilde{k}=1}^{\tilde{K}}  \sum_{j=1}^J -\text{log} \tilde{p}_{\tilde{k}j}
\right]\right) \\
&\text{subject to }\sum_{j=1}^J p_{kj} = 1\text{ and }\sum_{j=1}^J \tilde{p}_{\tilde{k} j} = 1 \label{sum_to_one} \text{ for all } k, \tilde{k} \text{ s.t. } \mathbf{p}_k, \tilde{\mathbf{p}}_{\tilde{k}} \text{ unknown }
\end{align}
and set $t^{(r + 1)} = at^{(r)}$ where $a>1$ is a prespecified incrementing factor, iterating until $t^{(r)} > t_{\text{cutoff}}$ for large $t_{\text{cutoff}}$. In practice we find that $t^{(0)} = 1$, $a = 10$, and $t_{\text{cutoff}} = 10^{12}$ yield good performance. We enforce the sum-to-one constraints (\ref{sum_to_one}) by reparametrizing $\mathbf{p}$ and $\tilde{\mathbf{p}}$ as $\boldsymbol{\rho}$ and $\tilde{\boldsymbol{\rho}}$, with
$\rho_{kj} := \text{log} \frac{p_{kj}}{p_{kJ}}$ and
$\tilde{\rho}_{\tilde{k}j} :=  \text{log} \frac{\tilde{p}_{\tilde{k}j}}{\tilde{p}_{\tilde{k}J}}$
for $j = 1, \dots, J - 1$, which
are well-defined because of the logarithmic penalty terms in (\ref{orig_par_penalty}).

In the second step of our optimization procedure,
we apply a constrained Newton algorithm within an augmented Lagrangian algorithm
to allow elements of $\hat{\mathbf{p}}$ and $\hat{\tilde{\mathbf{p}}}$ to equal zero.
 Iteratively in each row $\mathbf{p}_{k}$ of $\mathbf{p}$, we approximately solve
\begin{align}
&\text{arg min}_{\mathbf{p}_k}  f_n(\mathbf{p}_k)
\hspace{.25cm}\text{ subject to } \sum_{j = 1}^J p_{kj} = 1,~ p_{kj} \geq 0 \text{ for } j = 1, \dots J \label{profile_obj}
\end{align}
where we write $f_n(\mathbf{p}_k)$ as a function of only $\mathbf{p}_{k}$ to reflect that we fix all other parameters at values obtained in previous optimization steps. We choose update directions for $\mathbf{p}_k$
via an augmented Lagrangian algorithm of \citet{bazaraa2006nonlinear} applied to
\begin{align}
\mathcal{L}_k := Q_k^{(t)}+ \nu\left[\sum_{j = 1}^J p_{kj} - 1\right] + \mu\left[\sum_{j = 1}^J p_{kj} - 1\right]^2
\end{align}
where $Q_k^{(t)}$ is a quadratic approximation to $f_n(\mathbf{p}_k)$ at $\mathbf{p}_k^{(t)}$ and $\nu$ and $\mu$ are chosen using the algorithm of \citet{bazaraa2006nonlinear}.
The augmented Lagrangian algorithm iteratively updates $\nu$ and $\mu$ until solutions to $\mathcal{L}_k$ satisfy $|\sum_{j = 1}^J p_{kj} - 1| < \epsilon$ for a small
prespecified value of $\epsilon$ (we use $10^{-10}$ by default). Within each iteration of the augmented Lagrangian algorithm,
we minimize $\mathcal{L}_k$ via fast non-negative least squares to preserve nonnegativity of $\mathbf{p}_k$.
Through the augmented Lagrangian algorithm, we obtain
a value $\mathbf{p}_{\mathcal{L}_k}^{(t)}$ of $\mathbf{p}_k$ that minimizes $\mathcal{L}_k^{(t)}$ (at final values of $\nu$ and $\mu$) subject to nonnegativity constraints.
Our update direction for $\mathbf{p}_k$ is then given by $\mathbf{s}_k^{(t)} = \mathbf{p}^{(t)}_{\mathcal{L}_k} - \mathbf{p}^{(t)}_k$.
We conduct a backtracking line search in direction $\mathbf{s}_k^{(t)}$ to find an update $\mathbf{p}_k^{(t + 1)}$ that
 decreases $f_n(\mathbf{p}_k)$.



\section{Inference for $\mathbf{p}$ and $\boldsymbol{\beta}$} \label{sec:testing}

We now address construction of confidence intervals and hypothesis tests. We focus on parameters $\boldsymbol{\beta}$ and $\mathbf{p}$, which we believe to be the most common targets for inference.
To derive both marginal confidence intervals and more complex hypothesis tests, we
consider a general setting in which we observe some estimate $\hat{\phi} = \phi(\mathbb{P}_n)$ of
population quantity $\phi = \phi(P)$, where $\mathbb{P}_n$ is the empirical distribution
corresponding to a sample $\left\{\left(\mathbf{W}_i,\mathbf{Z}_i, \mathbf{X}_i, \tilde{\mathbf{Z}}_i\right)\right\}_{i = 1}^n$, $P$ is its population analogue, and $\phi$ is a Hadamard directionally
differentiable map into the parameter space $\Theta$ or into $\mathbb{R}$. To derive marginal confidence
intervals for $\mathbf{p}$ and $\boldsymbol{\beta}$, we let
$\phi(\mathbb{P}_n) = \boldsymbol{\hat{\theta}}_n$ with $\phi(P) = \boldsymbol{\theta}_0$. For
hypothesis tests involving multiple parameters, we specify $\phi(\mathbb{P}_n)$ as
$2\left[ \text{sup}_{\boldsymbol{\theta} \in \Theta} M_n(\boldsymbol{\theta}) - \text{sup}_{\boldsymbol{\theta} \in \Theta_0}M_n(\boldsymbol{\theta})\right]$
with population analogue $\phi(P) = 0$ under $H_0: \boldsymbol{\theta} \in \Theta_0$.
In each case, we estimate the asymptotic distribution of
$a(n)\left(\phi(\mathbb{P}_n) - \phi(P)\right)$ for an appropriately chosen $a(n) \rightarrow \infty$.

As our model
includes parameters that may
lie on the boundary of the parameter space,
 the limiting distributions of our estimators and test statistics in general do not have a simple distributional form
  \citep{geyer1994asymptotics},
 and the multinomial bootstrap
will fail to produce asymptotically valid inference \citep{andrews2000inconsistency}.
To address this, we employ
a Bayesian subsampled bootstrap \citep{ishwaran2009alternative},
which consistently estimates the asymptotic distribution of our estimators when the true parameter
is on the boundary. Let $\mathbbm{P}^{\boldsymbol{\xi}}_n$ be a weighted empirical distribution
$\sum_{i = 1}^n \xi_{i,n} \mathbf{1}_{\mathbf{W}_i}$ with weights
$\boldsymbol{\xi} \sim G$ for $G \sim \text{Dirichlet} \left(\frac{m}{n}\mathbf{1}_n\right)$. Then the bootstrap estimator
$a(m)\left(\phi\left(\mathbbm{P}_n^{\boldsymbol{\xi}}\right) - \phi\left(\mathbbm{P}_n\right)\right)$ converges weakly to the limiting distribution of $a(n)\left(\phi\left(\mathbbm{P}_n\right) - \phi\left(P\right)\right)$ if we choose $m = m(n)$ such that
$\underset{n \rightarrow \infty}{\text{lim}} m = \infty$ and $\underset{n \rightarrow \infty}{\text{lim}} \frac{m}{n} = 0$  \citep{ishwaran2009alternative}.
We explore finite-sample behavior of the proposed bootstrap estimators with $m = \sqrt{n}$ in Section \ref{simulations}, finding good Type 1 error control.

To derive marginal confidence intervals for elements of $\boldsymbol{\theta}$, we let $a(n) = \sqrt{n}$
and $\phi(P) = \boldsymbol{\theta}(P)$.
Then
 $\sqrt{m}(\boldsymbol{\hat{\theta}}^{\xi}_n -  \boldsymbol{\hat{\theta}}_n)$ has the same limiting distribution as $\sqrt{n}(\boldsymbol{\hat{\theta}} - \boldsymbol{\theta})$. Therefore, for $\hat{L}^{q}_{c}$ the $c$-th bootstrap quantile of the $q$-th element of $\sqrt{m}(\boldsymbol{\hat{\theta}}^{\xi}_n -  \boldsymbol{\hat{\theta}}_n)$
and $\hat{\theta}_q$ the $q$-th
element of the maximum (weighted) likelihood estimate $\boldsymbol{\hat{\theta}}$,
an asymptotically $100(1 - \alpha)\%$ marginal confidence interval for $\theta_q$ is given by $\left(\hat{\theta}_q - \frac{1}{\sqrt{n}} \hat{L}^{q}_{1 - \alpha/2}, ~ \hat{\theta}_q - \frac{1}{\sqrt{n}} \hat{L}^{q}_{\alpha/2} \right)$.

As it may be of interest to test hypotheses about multiple parameters while leaving other parameters unrestricted (e.g., $\boldsymbol{\beta} = \mathbf{0}$ with unrestricted elements of $\mathbf{p}$),
we also develop a procedure to test hypotheses of the form
$H_0: \{ \theta_k = c_k: k \in \mathcal{K}_0\}$ for a set of parameter indices $\mathcal{K}_0$ against alternatives with $\boldsymbol{\theta}$ unrestricted.
Letting $\Theta_0$ indicate the parameter space under $H_0$ and $\Theta$ indicate the full parameter
space, we conduct tests using test statistic
$T_n:= nT(\mathbbm{P}_n) := 2n[\text{sup}_{\boldsymbol{\theta} \in \Theta} M_n(\boldsymbol{\theta}) -  \text{sup}_{\boldsymbol{\theta} \in \Theta_0} M_n(\boldsymbol{\theta})]$.
As noted above, $T_n$ is in general not asymptotically $\chi^2$ if (unknown) elements of $\mathbf{p}$ or $\tilde{\mathbf{p}}$ lie
at the boundary, and so we instead approximate the null distribution of $T_n$ by bootstrap resampling from an empirical distribution
projected onto an approximate null; this is closely related to the approach suggested by \citet{hinkley1988bootstrap}.
Let $\mathring{\mu}_{ij}$ denote $\text{exp}(-\gamma_i)$ times the expectation of $W_{ij}$ under the full model; $\mathring{\mu}_{ij}^0$ denote
the analogous quantity under $H_0$; and define
$W_{ij}^0 = W_{ij} \frac{\mathring{\mu}_{ij}^0}{\mathring{\mu}_{ij}}$ $\text{ if } \mathring{\mu}_{ij} >0$ and otherwise set $W_{ij}^0 = W_{ij} =0$.
In practice, we do not know $\mathring{\mu}_{ij}$ or $\mathring{\mu}_{ij}^0$, so we replace $W_{ij}^0$ with
$\hat{W}_{ij}^0  = \frac{\hat{\mathring{\mu}}_{ij}^0}{\hat{\mathring{\mu}}_{ij}}$,
where $\hat{\mathring{\mu}}_{ij}$ and $\hat{\mathring{\mu}}_{ij}^0$ are, up to proportionality constant
$\text{exp}(\hat{\gamma}_i)$, fitted means for $W_{ij}$ under the full and null models.
 After constructing $\hat{\mathbf{W}}^0$,
we rescale its rows so row sums of $\hat{\mathbf{W}}^0$ and $\mathbf{W}$ are equal. We then approximate the null distribution of $T$ via bootstrap
draws from $mT(\mathbbm{P}^{\boldsymbol{\xi}}_{0n})$ where $\mathbbm{P}^{\boldsymbol{\xi}}_{0n}$ is the Bayesian subsampled bootstrap
distribution on $\hat{\mathbf{W}}^0_n$. We reject $H_0$ at level $\alpha$ if the observed likelihood ratio test statistic is larger than the $1 - \alpha$
quantile of the bootstrap estimate of its null distribution, or equivalently, if $T_n \geq \tilde{L}^m_{1 - \alpha}$ for $\tilde{L}^m_{1 - \alpha}$
the $1 - \alpha$ quantile of $mT(\mathbbm{P}^{\boldsymbol{\xi}}_{0n})$.

\section{Data Examples}

\subsection{Comparing detection effects across experiments} \label{subsec:costea}

We now demonstrate the utility of our model in comparing different experimental protocols for high-throughput sequencing of microbial communities. We consider data generated in the Phase 2 experiment of \citet{Costea:2017uv} (see also  \citet{McLaren:2019cn}), wherein ten human fecal specimens (labeled $1, 2, \ldots, 8$, A and B) were mixed with a synthetic community of 10 taxa and prepared for shotgun metagenomic sequencing according to three different sample preparations (labeled H, Q, and W;  samples A and B were only analyzed with preparation Q). The synthetic community was also sequenced alone. Raw sequencing data was processed into taxon abundance data using MetaPhlAn2 \citep{truong2015MetaPhlAn2} by \citet{McLaren:2019cn}.
 In addition to sequencing data, taxon abundances in the synthetic community were also measured using flow cytometry.
We treat both sequencing and flow cytometry measurements as outcomes $\mathbf{W}$.
We are interested in comparing the detection of taxa in the synthetic community across protocols H, Q and W relative to flow cytometry. We are specifically interested in testing the null hypothesis that all sequencing protocols share the
same detection effects.
To accomplish this, we estimate the $3\times10$ matrix $\boldsymbol{\beta}$ (we set $\boldsymbol{\beta}_{\cdot 10} = \mathbf{0}_3$ to ensure identifiability; see SI Section 9.2 for proof of identifiability). Each row of $\boldsymbol{\beta}$ corresponds to a sequencing protocol, and each column corresponds to a taxon.
For details regarding the specification of $\mathbf{X}$ and other model parameters, see SI Section 13.1.
Under our model, $\text{exp}(\beta_{1j})$,
$\text{exp}(\beta_{1j} + \beta_{2j})$, and $\text{exp}(\beta_{1j} + \beta_{3j})$
give the degree of over- or under-detection of taxon $j$ relative to taxon 10 under protocols H,
Q, and W, respectively. We compare this model
to a submodel in which $\beta_{kj} = 0$ for $k  = 2, 3$ and all $j$.
Under this null hypothesis, taxon detections relative to flow cytometry do not differ across protocols.

To compare predictive performance of these models, we perform 10-fold cross-validation
on each model (see SI Section 13.2 for details). 
We use a bootstrapped likelihood ratio test to
 formally test our full model against the null submodel. We use the Bayesian subsampled bootstrap with $m = \sqrt{n}$ to illustrate
its applicability, however, a multinomial bootstrap would also be appropriate as all parameters are in the interior of the parameter space in this case.  In addition, we report point estimates and
bootstrapped marginal 95\% confidence intervals for detection effects estimated for each protocol
under the full model. We also compare our results to MetaPhlAn2's ``plug-in'' estimate of each sample's composition.

\begin{figure}
\begin{centering}
\includegraphics[width=\textwidth]{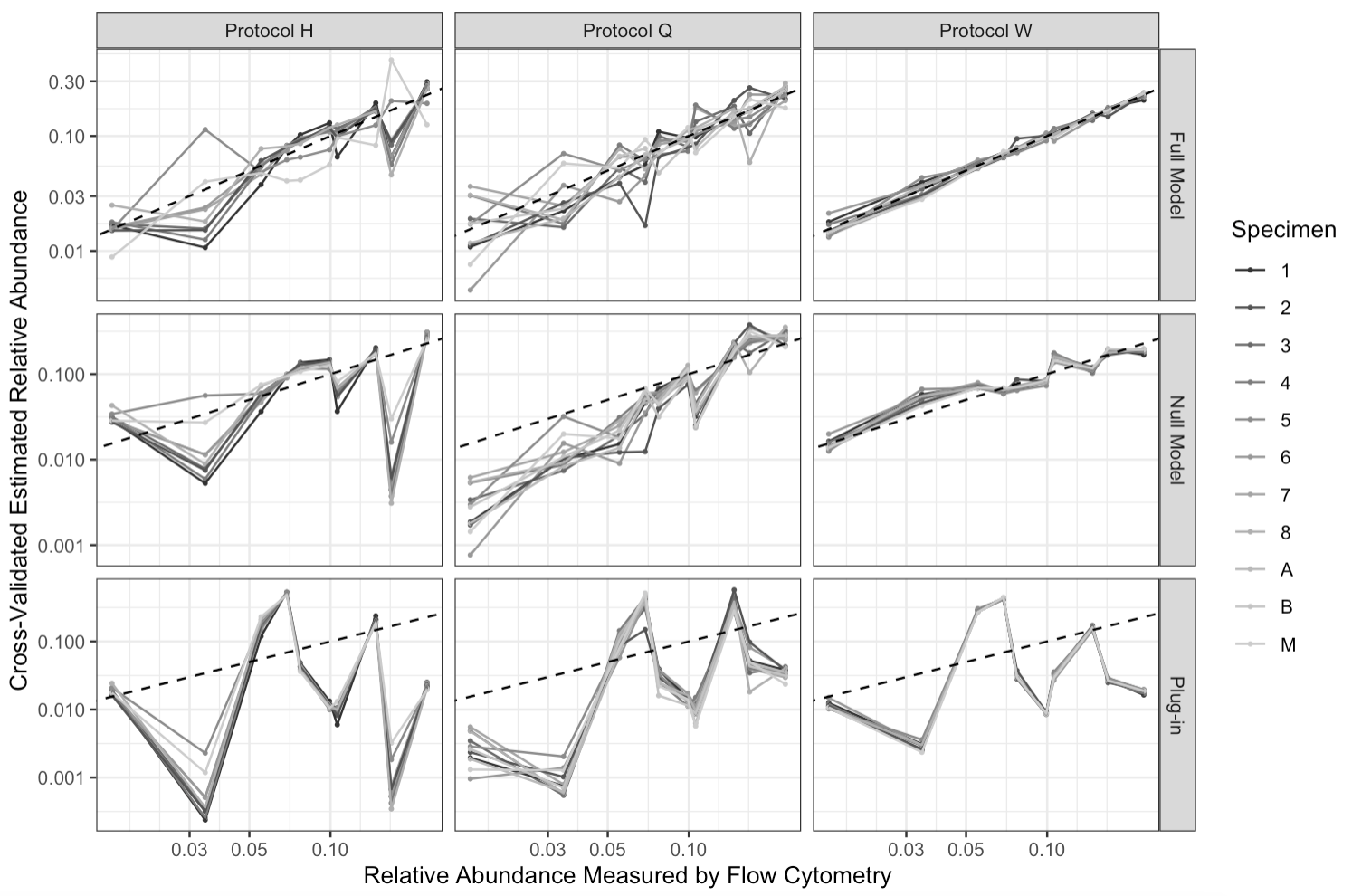}
 \caption{10-fold cross-validated estimates of relative abundance (y-axis) in \citet{Costea:2017uv} samples that were measured by
 whole-genome sequencing as well as plug-in estimates of relative abundance (bottom row).
 On the x-axis are relative abundance estimates obtained via flow cytometry (mean concentration in each taxon
 divided by sum across taxa). The top row contains estimates produced by
 a model containing separate detection effects $\boldsymbol{\beta}$ for each protocol; the middle row contains output from a model
 assuming a single common detection effect across protocols; and the bottom row contains ``plug-in'' estimates from
 MetaPhlAn2 output. Results for each protocols H, Q, and W are given in the leftmost, center, and rightmost columns, respectively.
 Within each pane, estimated relative abundances for the same sample are connected by line segments, and the line $y = x$ is indicated with a dotted line.} \label{costea_cv}
\end{centering}
\end{figure}

Figure \ref{costea_cv} summarizes 10-fold cross-validated estimated relative abundances
from the full model, null model and plug-in estimates.  We observe substantially better model fit for the full model (top row) than for the null model (middle row).
At each flow cytometric relative abundance, cross-validated estimates from the full model are generally centered around the line $y = x$ (dotted line),
whereas estimates from the null model exhibit substantial bias for some taxa. A bootstrapped likelihood ratio
test of the null model (i.e., $H_0: \boldsymbol{\beta}_{2\cdot} = \boldsymbol{\beta}_{3\cdot} = \mathbf{0}$) against the full model
reflects this, and we reject the null with $p < 0.001$.
Both the full and null models outperform
the plug-in estimates of sample composition (bottom row), which produces substantially biased estimates of
relative abundance relative to a flow cytometry standard. We report point estimates and marginal 95\% confidence intervals for $\boldsymbol{\beta}$ in SI Section 13.3.

Using the full model, we
estimate relative abundances with substantially greater precision under protocol W (top right) than under either other
protocol (top left and center). This appears to be primarily due to lower variability in measurements taken via
protocol W (bottom row). Our finding of greater precision of protocol W contrasts with \citet{Costea:2017uv}, who recommend protocol Q as a ``potential benchmark for new methods'' on the basis of
 median absolute error of
centered-log-ratio-transformed plug-in estimates of relative abundance against flow cytometry measurements (as well as on the basis of cross-laboratory measurement reproducibility, which we do not examine here).
The recommendations of \citet{Costea:2017uv} are driven by performance of plug-in estimators subject
to considerable bias, whereas we are able to model and remove a large degree
of bias and can hence focus on residual variation after bias correction.
We also note that \citet{Costea:2017uv} did not use MetaPhlAn2 to construct abundance estimates, which may partly account our different conclusions.

\subsection{Estimating contamination via dilution series} \label{subsec:karstens}

We next illustrate how to use our model to estimate and remove contamination in samples. We consider 16S rRNA sequencing data from \citet{karstens2019controlling}, who generated 9 samples via three-fold dilutions of a synthetic community containing 8 distinct strains of bacteria which each account for 12.5\% of the DNA in the community. Despite only 8 strains being present in the synthetic community, 248 total strains were identified based on sequencing (see SI Section 14.1 for data processing details). We refer to the 8 strains in the synthetic community as ``target" taxa and other strains as ``off-target."
Note that \citet{karstens2019controlling} identified one strain as a likely mutant of synthetic community member \textit{S. enterica}, and we refer to this strain as \textit{S. [unclassified]}.


To evaluate the performance of our model, we perform three-fold cross-validation and estimate relative abundance in the hold-out fold. We consider $\mathbf{p} \in \mathbb{R}^{2 \times 248}$, where the first row contains the known composition of the training fold $\left( \mathbf{0}^T_{240} ~\frac{1}{8}\mathbf{1}_8^T\right)$ and the second row is unknown and the target of inference (see SI Section 14.2 for full model specification). We evenly balance dilutions across folds, 
and using our proposed reweighted estimator to fit a model that accounts for the serial dilutions.
We set $\tilde{K}=1$ and let $\tilde{\mathbf{Z}}_i =  3^{d_i}$, where $d_i$ is the number of three-fold dilutions sample $i$ has undergone. 
This model reflects the assumption that the ratio of expected contaminant reads to expected non-contaminant reads is proportional to $3^{d_i}$. 
To avoid improperly sharing information about contamination amounts across folds,
we include terms in a fixed, unknown parameter
$\alpha\in \mathbb{R}$ in $\mathbf{D}_{\boldsymbol{\gamma},\boldsymbol{\alpha}}$. In particular, we let design matrix $\mathbf{V}$ by which we premultiply $\alpha$ consist of a single column vector with $i$-th entry 1 if sample
$i$ is in the held-out fold and 0 otherwise. This preserves information about relative dilution within the held-out fold without treating
samples in the training and held-out folds as part of the same dilution series. That is, the source of contamination for the held-out fold is modeled to 
be the same as in the training fold, but intensity of contamination for held-out sample $i$ having undergone $d_i$ three-fold dilutions is given by $\exp(\tilde{\gamma} + \alpha)3^{d_i}$ as compared to 
$\exp(\tilde{\gamma} )3^{d_i}$ for a sample in the training set. 
We model a single differential detection effect $\beta_j$ for each of the 8 taxa in the synthetic community, setting $\beta_{J} = 0$ for the reference taxon \textit{L. fermentum} for identifiability. Because $\beta_j$ is not identifiable for off-target taxa, we also fix $\beta_j = 0$ for $j=1, \ldots, 240$ (see SI Section 9.3 for identifiability results).

\begin{figure}
\begin{centering}
\includegraphics[width=\textwidth]{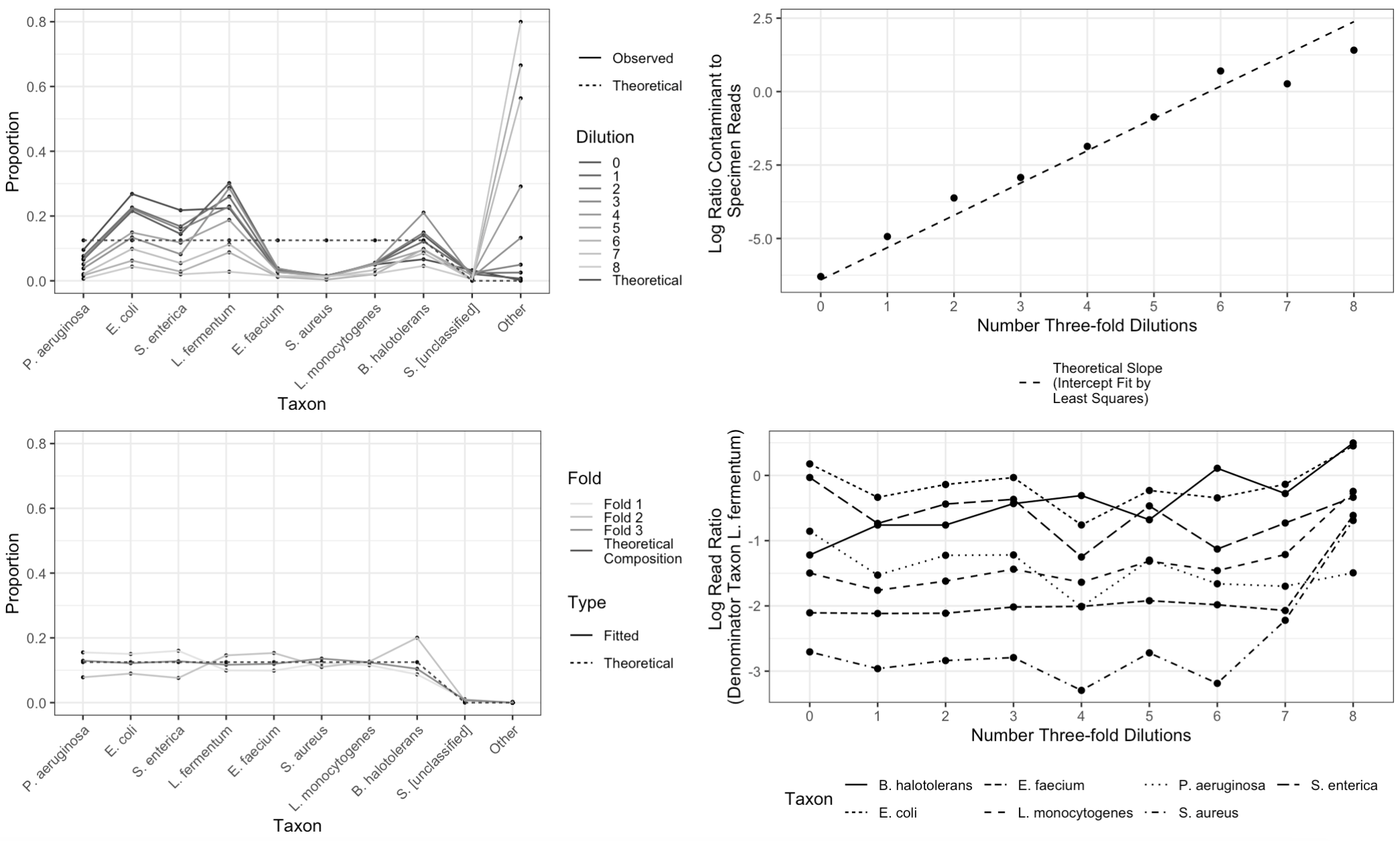}
\caption{Data from \citet{karstens2019controlling}, and estimates and summaries from the fitted model. (Top left) Observed read proportions by taxon and dilution, with theoretical synthetic composition
indicated by dotted line. (Top right) The log-ratio of total contaminant reads to total non-contaminant reads
(excluding \textit{S. [unclassified]}). The dotted line has slope
$\text{log}(3)$, with intercept fit via least squares. (Bottom left) Fitted read proportions obtained from each cross-validation fold,
with theoretical synthetic composition indicated by dotted line. Every fold produces abundance estimates that improve over observed read proportions. (Bottom right) The log-ratio of reads for target taxa to reference taxon \textit{L. fermentum} is relatively constant across increasing numbers of dilutions.
}  \label{karstens_1}
\end{centering}
\end{figure}

Figure \ref{karstens_1} shows data from \citet{karstens2019controlling} along with summaries  of our analysis.
Our estimate for the relative abundance of taxa in samples in the held-out fold $\mathbf{p}_{2 \cdot}$  improves on the performance of plug-in estimators (Figure \ref{karstens_1}, left panels) by taking into account two forms of structure in the \citet{karstens2019controlling} data. First, in each
successive three-fold dilution, we observe approximately three times more contamination relative to the
number of non-contaminant reads (Figure \ref{karstens_1}, top right). In addition, our model accounts for the degree of under- (or over-) detection of target taxa relative to \textit{L. fermentum}. We observe that taxon detection is reasonably constant across dilutions (Figure \ref{karstens_1}, bottom right). However, we do observe greater variability in taxon detections at higher dilutions, most likely because we observe comparatively few reads ($\sum_{j=1}^{248} W_{\{i: d_i = 1\}j} \approx 227,000$ while $\sum_{j=1}^{248} W_{\{i: d_i = 8\}j} \approx 8,000$).

In terms of root mean squared error (RMSE) $\sqrt{1/J\sum_{j=1}^J (\hat{p}_j - p_j)^2}$, our cross-validated estimates ($0.0037$, $0.0073$, and $0.0033$) substantially outperform
 the ``plug-in'' estimates given
by sample read proportions in any of these dilutions (median $0.017$; range $0.013$ -- $0.022$). This
is not an artifact of incorporating information from 3 samples in each cross-validation fold, as pooling reads across all samples yields an estimator with RMSE $0.014$.

Fitting a model to this relatively small dataset and evaluating its performance using cross validation prohibits the reasonable construction of confidence intervals. Therefore, to evaluate the performance of our proposed approach to generating confidence intervals, we also
 fit a model which treats all samples as originating from a single specimen of unknown composition. $\boldsymbol{\beta}$  is not identifiable in this setting, so we set it equal to zero and do not estimate it. We set $\tilde{K}=1$ and $\tilde{\mathbf{Z}}_i = \text{exp}(\gamma_i) \times 3^{d_i}$ as before (the need for $\tilde{\alpha}$ is alleviated). Strikingly, under this model, marginal 95\% confidence intervals for elements of $\mathbf{p}$ included zero for 238 out of 240 off-target taxa (empirical coverage of $99.2\%$ when true $p_{kj} = 0$).
No interval estimates for target taxa included zero.
This suggests that 
applying our proposed approach to data from a dilution series experiment can aid in evaluating whether taxa detected by next-generation sequencing are actually present in a given specimen.


\section{Simulations} \label{simulations}

\subsection{Sample size and predictive performance}

To evaluate the predictive performance of our model on microbiome datasets
of varying size, we use
data from 65 unique specimens consisting of synthetic communities of 1, 2, 3, 4, or 7 species combined in equal abundances.  \citet{brooks2015truth} performed 16S amplicon sequencing on 80 samples of these 65 specimens across two plates of 40 samples each. Very few reads in this dataset were ascribed
to taxa outside the 7 present by design, so we limit analysis to
these taxa. To explore how prediction error varies with
number of samples of known composition, we fit models treating randomly selected
subsets of $n_{\text{known}} \in \{3, 5, 10, 20\}$ samples per plate as known.
In each model, we included one source of unknown contamination for each plate and estimate a detection vector $\boldsymbol{\beta} \in \mathbb{R}^{J}$. For each $n_{\text{known}}$, we drew
100 independent sets of samples to be treated as known, requiring that each set satisfy an identifiability condition
in $\boldsymbol{\beta}$ (see SI Section 9.4).
On each set, we fit both the unweighted and reweighted
models, treating $n_{\text{known}}$ samples as having known composition and $80-n_{\text{known}}$ samples as arising from unique
specimens of unknown composition.

We observe similar RMSE for the reweighted and unweighted estimators. For $n_{\text{known}}$ equal to
3, 5, 10, and 20, we observe RMSE 0.041, 0.037, 0.035, and 0.032 (to 2 decimal places) for both estimators.
 By comparison, the RMSE for the plugin estimator $\frac{W_{ij}}{\sum_{j=1}^J W_{ij}}$
 is 0.173. Notably, RMSE decreases but does not approach zero as larger
 number of samples are treated as known, which  reflects that we estimated each
 relative abundance profile on the basis of a single sample. 

With respect to correctly estimating $p_{kj}$ when $p_{kj}=0$, we again see very similar performance of the reweighted and unweighted Poisson
estimators. Out of 100 sets, unweighted
estimation yields $\hat{p}_{kj}=0$ for $53\%, 55\%, 59\%$, and $64\%$ of $\{k, j\}$ pairs for which $p_{kj}=0$ ($n_{\text{known}}$ equal to
3, 5, 10, and 20). The corresponding figures for the reweighted estimator
are $53\%, 55\%, 58\%$, and $63\%$. The plug-in estimator sets $51\%$ of these
relative abundances equal to zero.
While we observe the proportion
of theoretical zero relative abundances estimated to be zero increases in
number of samples treated as known regardless of estimator, in general
we do not expect this proportion to approach 1 as number of known samples
increases.
We also note that our model is not designed to
 produce prediction intervals, and confidence intervals for a parameter estimated from a
single observation are unlikely to have reasonable coverage. Finally, we acknowledge that despite the excellent performance of our model on the data of \citet{brooks2015truth}, our model does not fully capture sample cross-contamination
known as index-hopping \citep{hornung2019issues}, which likely affects this data.


\subsection{Type 1 error rate and power}

To investigate the Type 1 error rate and power of tests based on reweighted and unweighted estimators, we simulate data arising from a set of hypothetical dilution series. In each simulated dataset, we observe reads from dilution series of four specimens: two specimens of known composition and two specimens of unknown composition (specimens A and B). 
Each dilution series consists of four samples:
an undiluted sample from a specimen as well as a 9-, 81-, and 729-fold dilution of the specimen.
We vary the number of taxa $J \in \{5, 20\}$, as well as the magnitude of elements of $\boldsymbol{\beta}$,  the number of samples, and the distribution of $W_{ij} | \mu_{ij}$. The identifiability results of SI Section 9.3 apply here. 

We consider three different values of $\boldsymbol{\beta} \in \mathbb{R}^{J}$: $\boldsymbol{\beta} = 0 \boldsymbol{\beta}^{\star}$,
$\boldsymbol{\beta} = \frac{1}{10} \boldsymbol{\beta}^{\star}$,
and  $\boldsymbol{\beta} = \boldsymbol{\beta}^{\star}$ where $\boldsymbol{\beta}^{\star} = \begin{pmatrix} 3& -1 & 1 & -3 & 0 \end{pmatrix}$
when $J = 5$ and $ \boldsymbol{\beta}^{\star} = \begin{pmatrix} 3& -1 & 1 & -3 & 3 & -1 & \dots & -3 & 0 \end{pmatrix}$ when $J = 20$.
We base the magnitude of entries of $\boldsymbol{\beta}^{\star}$ using the observed magnitude
of entries of $\boldsymbol{\hat{\beta}}$ in our analysis of \citet{Costea:2017uv} data. We vary the number of samples between either a single dilution series from each specimen or three dilution series from each specimen (for a total of nine samples per specimen). We draw $W_{ij}|\mu_{ij}$ from either a Poisson($\mu_{ij}$) distribution or a Negative Binomial
distribution with mean parameter $\mu_{ij}$ and size parameter $s = 13$. $s=13$ was chosen to approximate the \citet{karstens2019controlling} data via a linear regression of fitted mean-centered squared residuals.
In all settings we simulate $\{\gamma_i\}_{i = 1}^n$ from a log-normal distribution with parameters $\mu = \text{min}(13.5 - 1.5d_i,12)$ and
$\sigma^2 = 0.05$. These values were chosen based on observed trends in reads from target taxa in the data of \citet{karstens2019controlling} data.
In all settings, the first specimen has true relative abundance proportional to $(1, 2^{\frac{4}{J - 1}}, 2^{2\cdot \frac{4}{J - 1}}, \dots, 2^4)$, that is,  taxon $J$ is 16 times more abundant than taxon 1. The second specimen has true relative abundance proportional to $(2^4, \dots, 2^{\frac{4}{J - 1}}, 1)$. When $J = 5$, the first two taxa are absent from specimen A, and when $J = 20$, first eight taxa are absent from specimen A. Relative abundances in the remaining
taxa form an increasing power series such that the first taxon present in nonzero abundance
has relative abundance that is $1/100$-th of the relative abundance of taxon $J$. The relative abundance profile of specimen
B is given by the relative abundance vector for specimen A in reverse order. We also simulate the degree of contamination as scaling with dilution. When comparing samples with the same read depth, on
average a 9-fold diluted sample will contain 9 times more contaminant reads than an undiluted sample (see Section 5.2 and Figure \ref{karstens_1}, top right).
We simulate contamination from a source containing equal relative abundance of all taxa.
We set $\tilde{\mathbf{Z}} \in \mathbb{R}^{n}$ such that $\tilde{Z}_i = 9^{d_i}$ where $d_i$ is the degree of dilutions of sample $i$, and $\tilde{\gamma} = -3.7$,
as we observe in \cite{karstens2019controlling} data.

\begin{figure}
\begin{centering}
\includegraphics[width=\textwidth]{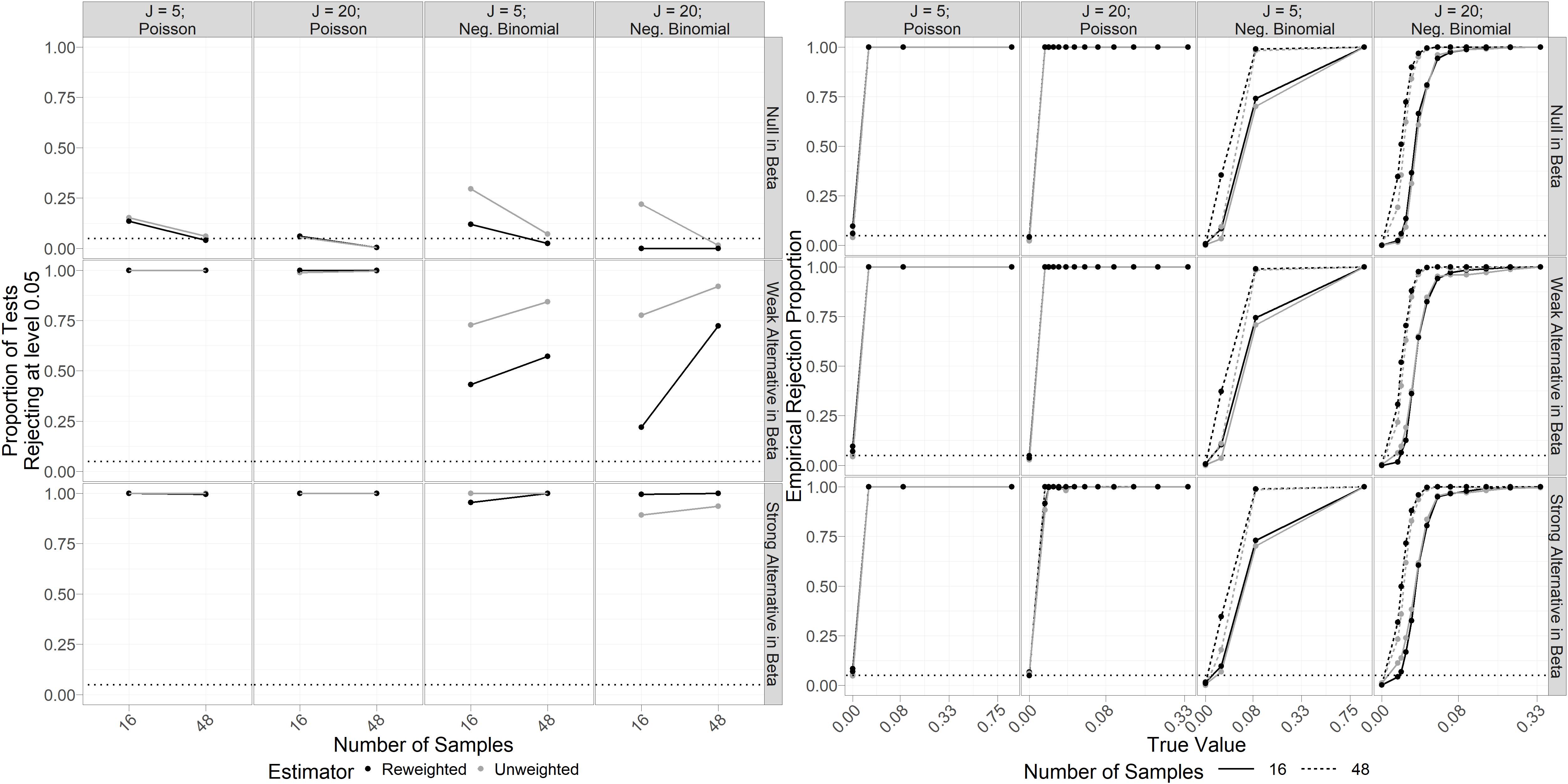}
\caption{At left, the Type 1 error (top row) and power of our proposed likelihood ratio tests for both the unweighted and reweighted estimators.
Performance of tests of $H_0: \boldsymbol{\beta} =0$ against
a general alternative are summarized in terms of empirical rejection rates at level $\alpha = 0.05$ (y-axis), with sample size plotted
on the x-axis. Columns give the conditional distribution of data (Poisson or Negative Binomial) and number of taxa $J$. Rows specify
whether data was simulated under the null $\boldsymbol{\beta} = 0$, under a ``weak'' alternative with $\boldsymbol{\beta} = \frac{1}{10} \boldsymbol{\beta}^{\star}$ (i.e., $\boldsymbol{\beta} \neq 0$ of small magnitude),
or under a ``strong'' alternative $\boldsymbol{\beta} = \boldsymbol{\beta}^{\star}$ (i.e., $\boldsymbol{\beta} \neq 0$ of larger magnitude).
At right, empirical rejection rate for marginal bootstrap tests of $H_0: p_{kj} = 0$ at the $0.05$ level
versus true value of $p_{kj}$ (x-axis). Columns and rows are as specified above.}
\label{simulation_summary}
\end{centering}
\end{figure}

 Figure \ref{simulation_summary} summarizes our results from 250 simulations under each condition.
 Empirical performance of bootstrapped likelihood ratio tests of
$H_0: \boldsymbol{\beta} = 0$ against $H_A: \boldsymbol{\beta} \neq 0$ at level 0.05 (lefthand pane of Figure \ref{simulation_summary})
reveals good performance, even for small sample sizes. Unsurprisingly, we observe improved Type 1 error control and power at larger sample sizes.
Tests based on the reweighted estimator generally improved Type 1 error compared to the unweighted estimator, with the greatest improvements observed for data
simulated from a negative binomial distribution (mean Type 1 error across simulations was 0.15 for the unweighted estimator and 0.04 for the reweighted estimator).
Surprisingly, Type 1 error control appears to improve
when the number of taxa $J$ is larger (mean Type 1 error was 0.11 for $J =5$ and 0.05 for $J = 10$). This may in part be
a result of simulating $W_{ij}$ as conditionally independent
given $\gamma_i$, covariates, and other parameters.
Power to reject the null hypothesis is very high when the data generating process is Poisson, as well as for strong alternatives ($\boldsymbol{\beta}= \boldsymbol{\beta}^{\star}$)
when the data follows a negative binomial distribution.

Empirical Type 1 error control for bootstrapped marginal tests of $H_0: p_{kj} = 0$ against $H_A: p_{kj} >0$
 at the 0.05 level (lefthand pane
of Figure \ref{simulation_summary})
is generally no larger than nominal, with median empirical Type 1 error $0.02$ across all conditions.
We observe above-nominal Type 1 error in limited cases, primarily when sample size is small
and data is Poisson-distributed (the largest observed Type 1 error rate, of 0.10, occurs for
 tests based on reweighted estimators at sample size $n = 16$ and number of taxa $J = 5$
 when data is Poisson-distributed). Power to reject the null is close to 1 for
 all non-zero values of $p_{kj}$ when data is conditionally Poisson-distributed
 (median empirical power across conditions 1; minimum 0.88). When data is
 simulated as negative binomial, unsurprisingly, power appears to increase in magnitude
 of $p_{kj}$ for tests based on unweighted and reweighted estimators, with somewhat
 higher power in tests using reweighted estimators. Magnitude of $\boldsymbol{\beta}$
 (rows of Figure \ref{simulation_summary}) does not appear to affect Type 1 error
 or power of our tests of $H_0: p_{kj} = 0$. We also note that empirical coverage
 of marginal bootstrapped confidence intervals for $p_{kj}$ is generally lower
 than nominal in our simulations (SI Figure 2). This aligns with
  general performance of bootstrap percentile confidence intervals at small sample sizes,
  and we expect that generating confidence intervals from
  inverted bootstrapped likelihood ratio tests would yield better
  coverage in this case (unfortunately this is not  feasible for computational reasons).

\section{Discussion}


In this paper, we introduce a statistical method to model measurement error
due to contamination and differential detection of taxa in microbiome experiments. Our method builds on
previous work in several ways.
By directly modeling the output of
microbiome experiments, we do not rely on
data
transformations that discard information regarding measurement
precision, such as ratio- or proportion-based transformations. 
This affords our method the key advantage of estimating relative abundances
lying on the boundary of the simplex, which is typically precluded by transformation-based approaches. Accordingly, we implement inference tools appropriate to the non-standard parameter space that we consider.
The advantage of estimating relative abundances on the boundary of the simplex is not purely theoretical, and we show that our interval estimates do indeed include boundary values, and demonstrate above-nominal empirical coverage in an analysis of data from \citet{karstens2019controlling}.
Furthermore, our reweighting estimator allows for flexible mean-variance relationships without the need to specify a parametric model. 
Our approach to parameter estimation does not assume that observations are counts, and therefore our method can be applied to a wide array of microbiome data types, including proportions, coverages and cell concentrations as well as counts.
Finally, our method can accommodate complex experimental designs, including analysis of mixtures of samples, technical replicates, dilution series, detection effects that vary by experimental protocol or specimen type, and contamination impacting multiple samples. Contamination is commonly addressed via ``pre-processing'' data, thereby conditioning on the decontamination step.
In contrast, by simultaneously estimating contamination along with all other model parameters, our approach captures holistic uncertainty in estimation.


Another advantage of our methodology is that we do not require the true composition of any specimens to be known. For example, our
test of equal detection effects across protocols in \citet{Costea:2017uv} can be performed without
knowledge of specimen composition.
Accordingly, our approach provides a framework for comparing experimental protocols in the absence of synthetic communities, which can be challenging to construct. 

In addition, we expect that our method may have substantial utility
applied to dilution series experiments, as illustrated in our analysis of data from
\citet{karstens2019controlling}. Dilution series are relatively low-cost and scalable (especially in comparison to synthetic communities) and may be especially advantageous when the impact of sample contamination on relative abundance estimates is of particular concern.
With this said, we strongly recommend against testing the composite null $H_0:p_{kj} >0$ against the point alternative $H_A: p_{kj}=0$, as these hypotheses are statistically indistinguishable on
the basis of any finite sample of reads.
However, we are able to determine the degree to which
our observations are consistent with the \textit{absence} of any given taxon, and therefore we can
meaningfully test $H_0:p_{kj} = 0$ against a general alternative.

The focus of our paper was on $\mathbf{p}$ and $\boldsymbol{\beta}$ as targets of inference. Future research could investigate extensions to our model that connect relative abundances $\mathbf{p}$ to covariates of interest, allowing the comparison of average relative abundances across groups defined by covariates, for example.
Our proposed bootstrap procedures may also aid the propagation of uncertainty to group-level comparisons of relative abundances in downstream analyses. In addition, while we focus applications of our model on microbiome data, our model could be applied to a broad variety of data structures obtained from high-throughput sequencing, such as single-cell RNAseq. We leave these applications to future work.

\section{SI: Identifiability}

As discussed in the main text, our model is flexible enough to encompass a wide variety of experimental designs and targets of estimation, but as a result, the parameters in model (5) are unidentifiable without additional constraints. In this section, we demonstrate identifiability for the three use cases of our model that we considered as illustrative examples. For reference, we provide here the definition of identifiability:

\begin{defi} \label{ident_defn}
  A mean function $\mu: \mathcal{D} \subset \mathbb{R}^q \rightarrow \mathbb{R}^N$ is said to be \textit{identifiable} if $\theta \neq \theta' \Rightarrow \mu(\theta) \neq \mu(\theta')~ \forall ~\theta, \theta' \in \mathcal{D}.$
  \end{defi}

\subsection{Preliminaries}

We first consider situations where contamination is not modelled (e.g., Section 6.1), and as such, $\tilde{\mathbf{Z}} = 0$. 
In this case, the following Lemma provides a sufficient condition under which mean model (5) may be reparametrized as a log-linear model. 
Since a log-linear model is identifiable if 
and only if its design matrix has full column rank, this result provides a basis on which to establish identifiability for a large subclass of models. 

Unless otherwise noted, without loss of generality we take $\beta_{.J} = \mathbf{0}_p$ throughout. 
 
\begin{lem} \label{ident_lemma}
 Let $\mathbb{E}[\mathbf{W}|\boldsymbol{\beta}, \mathbf{p}, \boldsymbol{\gamma}]=: \boldsymbol{\mu}(\boldsymbol{\beta}, \mathbf{p}, \boldsymbol{\gamma})$ denote a mean function for 
 an $n \times J$ matrix $\mathbf{W}$ such that the $(i, j)$-th element of $\boldsymbol{\mu}(\boldsymbol{\beta}, \mathbf{p}, \boldsymbol{\gamma})$ is given by
 $\mu_{ij}(\boldsymbol{\theta}) =  \exp(\gamma_i)Z_ip^j\exp(\mathbf{1}_{\{j<J\}}X_i\beta^j)$,
for $\gamma_i$ the $i$-th element of $\boldsymbol{\gamma}\in \mathbb{R}^n$, 
$p^j$ the $j$-th column of $\mathbf{p}$ (for $\mathbf{p}$ a $K \times J$ matrix with rows in the $(J-1)$-simplex), 
$\beta^j$ the $j$-th column of the $p\times(J - 1)$ matrix $\boldsymbol{\beta}$, and
$Z_i$ and $X_i$ the $i$-th rows of $\mathbf{Z} \in \mathbb{R}_{\geq 0}^{n\times K}$ and $\mathbf{X} \in \mathbb{R}^{n\times p}$, respectively. 
If each row of $\mathbf{p}$ lies in the interior of $\mathbb{S}^{J - 1}$ and each row of $\mathbf{Z}$ contains a single nonzero entry $z_i$, 
 equal to 1, 
  then there exists a matrix $M \in \mathbb{R}^{nJ\times q}$ for $q = (p + K)(J - 1) + n$ and an isomorphism 
 $f: \mathbb{R}^{p\times (J - 1)} \times \mathbf{S}^{K\times (J - 1)} \times \mathbb{R}^{n} \rightarrow \mathbb{R}^{q}$ such that for all 
 $(\boldsymbol{\beta}, \mathbf{p}, \boldsymbol{\gamma})$,
$\mu_{ij}(\boldsymbol{\beta}, \mathbf{p}, \boldsymbol{\gamma}) = \exp\big(M_{[ij]} f(\boldsymbol{\beta}, \mathbf{p}, \boldsymbol{\gamma})\big)$ for $M_{[ij]}$ the $(i - 1)\times J + j$-th row of $M$. 
\end{lem}

\begin{proof} It suffices to show the result holds for an arbitrary element of $\boldsymbol{\mu}(\boldsymbol{\beta}, \mathbf{p}, \boldsymbol{\gamma})$, say, $\mu_{ij}$. Let $l$ be the unique index such that $z_{il} := 1$. We have 
\begin{align*}
\mu_{ij} & = \exp(\gamma_i)Z_ip^j\exp(\mathbf{1}_{\{j<J\}}X_i\beta^j) \\
&= z_{il}p_{lj}\exp(\gamma_i + \mathbf{1}_{\{j<J\}}X_i\beta^j) \\
& = \exp(\gamma_i + \log p_{lJ}  + \mathbf{1}_{\{j<J\}}X_i \beta^j + \log \frac{p_{lj}}{p_{lJ}}) \\
& =: \exp(\delta_i  + \mathbf{1}_{\{j<J\}}X_i \beta^j + \mathbf{1}_{\{j<J\}} \rho_{lj} )  \text{ for } \delta_i := \gamma_i + \log p_{lJ} \text{ and } \rho_{lj} :=  \log \frac{p_{lj}}{p_{lJ}} 
\end{align*}
We let $f(\boldsymbol{\beta}, \mathbf{p}, \boldsymbol{\gamma})  = [\delta_1, \dots,  \delta_n, \rho_{11}, \dots, \rho_{1(J - 1)}, \dots, \rho_{K1}, \dots, \rho_{K(J - 1)}, {\beta^1} ^T\dots, {\beta^{J - 1}}^T]^T$, which is 
trivially an isomorphism. $M_{[ij]} \in \mathbb{R}^{1 \times q}$ is then given by
\begin{align}
M_{[ij]} & = [e_i^T :  \mathbf{1}_{\{j<J\}}e_l^T \otimes e_j^T: \mathbf{1}_{\{j<J\}}e_j^T\otimes X_i] \label{eqn:m_form}
\end{align}
\noindent where $\otimes$ is the Kronecker product, $e_i \in \mathbb{R}^n$ has $i$-th element equal to 1 and all others 0, $e_l \in \mathbb{R}^K$ has
$l$-th element equal to 1 and all others 0, and 
$e_j \in \mathbb{R}^{J-1}$ has $j$-th element equal to 1 and all others zero (when $j=J$, $e_j$ contains only zeroes). Matrix multiplication then gives $\mu_{ij}(\boldsymbol{\beta}, \mathbf{p}, \boldsymbol{\gamma}) =\exp\big( M_{[ij]}f(\boldsymbol{\beta}, \mathbf{p}, \boldsymbol{\gamma})  \big)$.
\end{proof}

\begin{thm} \label{ident_iff_thm}
Under the conditions of Lemma \ref{ident_lemma}, then mean function $\mu$ is identifiable if and only if $M$ has full column rank.
\end{thm}
\begin{proof}
  We proceed by contradiction for both directions. ($\Leftarrow$) 
  Suppose $\mu$ satisfies the criteria given in Lemma \ref{ident_lemma} and is not identifiable. For $\mathbf{\theta} = (\boldsymbol{\beta}, \mathbf{p}, \boldsymbol{\gamma})$, this 
implies that $\mu(\theta) = \mu(\theta')$ for some $\theta, \theta'$ s.t. $\theta \neq \theta'$. 
This in turn implies that $M f(\theta)  = M f(\theta') \Rightarrow M \big[ f(\theta) - f(\theta') \big] = \mathbf{0} \Rightarrow f(\theta) - f(\theta')  = \mathbf{0}$ since $M$ has full column rank. Since $f$ is an isomorphism,
we have $\theta = \theta'$, and therefore, a contradiction. Hence $\mu$ must be identifiable if $M$ is full rank.
($\Rightarrow$) Now suppose $M$ is not full rank. Then for some $a \neq \mathbf{0} \in \mathbb{R}^q$, $Ma = \mathbf{0}$. Letting $\eta$ be an arbitrary vector in $\mathbb{R}^q$, we have $M\eta = M(\eta + a)$ and hence 
$\mu(f^{-1}(\eta)) = \mu(f^{-1}(\eta + a))$. Since $f$ is an isomorphism, we must have $f^{-1}(\eta) \neq f^{-1}(\eta + a)$, and hence the mean function is not identifiable.
\end{proof}

The above results apply when all rows of $\mathbf{p}$ are unknown. However, when some rows of $\mathbf{p}$ are known, we may establish identifiability of the unknown parameters under the conditions of Lemma \ref{ident_lemma} by considering the rank of the matrix $M'$ obtained from $M$ by excluding columns corresponding to any element of $\rho_k$ for $k$ such that $p_k$ is known.



\subsection{Identifiability: Comparing detection effects across experiments} \label{subsec:identifiability_costea}

In Section 6.1, we analyzed data consisting of measurements taken via three sequencing protocols on human fecal samples containing a spiked-in artificial community in addition
to measurements taken via flow cytometry on the spiked-in community \citep{Costea:2017uv}. We constrained our analysis to taxa in the spiked-in community, which by design are present in every sample. In this setting, we may apply 
Theorem \ref{ident_iff_thm}. 

We first show identifiability of the mean function defined for models fit without cross-validation, and then address models fit via cross-validation. 
We assume familiarity with the experimental setup and notation introduced in Section 6.1. 
We adopt the identifiability constraint $\beta^J = 0$. 

\subsubsection{Model without Cross-Validation}

In the model fit without cross-validation, all samples originate from a single source. The mean for sample $i$ is given by
\begin{align}
\boldsymbol{\mu}_i = \exp(\gamma_i)[\bold{p}\circ\exp(X_i\boldsymbol{\beta})]
\end{align}
where $\gamma_i$ parametrizes sample intensity, $\boldsymbol{\beta} \in \mathbb{R}^{p \times J}$ is a matrix of detection effects (here, $p=3$),
$\mathbf{p}$ is the unknown relative abundance profile for the sample from which our observations were taken, and  
$$X_i =\begin{cases}  \mathbf{0} &\text{ if sample } i \text{ was measured by flow cytometry,} \\  [1 ~\mathbf{1}_{Q} ~ \mathbf{1}_{W}] & \text{ otherwise}\end{cases}$$ for $\mathbf{1}_{Q} $ and $\mathbf{1}_{W}$ indicators
for sample $i$ being processed according to protocol $Q$ or $W$, respectively. Thus, from (\ref{eqn:m_form}), the form of $M$ is 
\begin{align*}
M_{[ij]} 
& =  \begin{cases} [e_i^T : e_j^T: \mathbf{0}^T] &\text{ if sample } i \text{ was measured by flow cytometry,} \\
[e_i^T : e_j^T: \mathbf{1}_{\{j<J\}}e_j^T\otimes [1 ~ \mathbf{1}_Q ~\mathbf{1}_W] ] & \text{ otherwise.}
\end{cases}
\end{align*}
Hence the submatrix of $M$ consisting of rows corresponding to taxa $1$ through $J$ in sample $i$ is given by
$M_i = [1_J e_i^T : I_{-J} : I_{-J} \otimes X_i]$
where $1_J$ is a column $J$-vector of ones and $I_{-J}$ is the $J\times J$ identity matrix with $J$-th column removed.
Since there are 4 unique rows of $X$, the leftmost $(p + 1)\times(J - 1) =  4\times(J - 1) $ columns of $M$ are spanned by the $4\times(J - 1)$ rows of
\begin{align*}
&\begin{bmatrix} 
I_{-J} &:&  I_{-J} \otimes [0~0~0] \\  I_{-J} &:& I_{-J} \otimes [1 ~0 ~0] \\  I_{-J}  &:& I_{-J} \otimes [1 ~1 ~0] \\  I_{-J} &:& I_{-J} \otimes [1 ~0~ 1] \\
\end{bmatrix}. 
\end{align*} 
Since this is a full-rank matrix, the matrix consisting of the leftmost $(p + 1)\times(J - 1) $ columns of $M$ must have full column rank. It now remains to show that the rightmost $n$ columns of $M$ are 
linearly independent of the leftmost  $(p + 1)\times(J - 1) $ and compose a matrix with full column rank. The latter is immediate from inspection, and the former follows from the fact that the leftmost  $(p + 1)\times(J - 1) $
columns of $M$ contain rows of zeroes (corresponding to observations on taxon $J$), whereas the rightmost $n$ columns contain no zero rows and hence must be linearly independent of the remainder of the columns. Theorem 1 therefore applies and we have established identifiability of this model.

\subsubsection{Cross-Validated Model}

The cross-validated version of the above model 
differs only in the addition of an extra row of $\mathbf{p}$ corresponding to the 
composition of the specimen from which samples in the held-out fold are taken. 
This yields a model we can express as a log-linear model with nearly the same design matrix 
as above. Namely, we have $M$ such that the the submatrix consisting of rows corresponding to sample $i$ is given by
\begin{align*}
M_i &= [1_J e_i^T : \mathbf{1}_{\{i \text{ held-in}\}} I_{-J} : \mathbf{1}_{\{i \text{ held-out}\}} I_{-J} : I_{-J} \otimes X_i]
\end{align*}
\noindent where $\mathbf{1}_{\{i \text{ held-in}\}}$ is an indicator that sample $i$ is in the held-in fold, and analogously for 
$\mathbf{1}_{\{i \text{ held-out}\}}$. 
There are now 8 rather than 4 unique rows in columns of $M$: 
\begin{align*}
&\begin{bmatrix} 
I_{-J} &:&\mathbf{0} &:&  I_{-J} \otimes [0~0~0] \\  I_{-J} &:&\mathbf{0}  &:& I_{-J} \otimes [1 ~0 ~0] \\  I_{-J} &:&\mathbf{0} &:& I_{-J} \otimes [1 ~1 ~0] \\  I_{-J} &:& \mathbf{0}&:& I_{-J} \otimes [1 ~0~ 1] \\
 \mathbf{0} &:& I_{-J} &:& I_{-J} \otimes [0~0~0] \\   \mathbf{0} &:& I_{-J} &:& I_{-J} \otimes [1 ~0 ~0] \\   \mathbf{0} &:& I_{-J} &:& I_{-J} \otimes [1 ~1 ~0] \\  \mathbf{0} &:& I_{-J} &:&  I_{-J} \otimes [1 ~0~ 1] 
\end{bmatrix} 
\end{align*} 
This is again a full-rank matrix, and thus the columns of $M$ corresponding to $\boldsymbol{\rho}$ and $\boldsymbol{\beta}$ are linearly independent. Consequently, Theorem 1 applies and thus the mean model is identifiable.

%
%
%

\subsection{Identifiability: Estimating contamination via dilution series} \label{subsec:identifiability_karstens}

In Section 6.2, we analyze a serial dilution of a specimen of known composition \citep{karstens2019controlling}. In this application, we estimated a contamination profile, and therefore the conditions of Theorem \ref{ident_iff_thm} 
do not apply. In the following sections we thus establish identifiability of the model directly. We consider the two cases discussed in the manuscript (fitting with and without cross-validation) separately. 
 
\subsubsection{Two sources; estimated via cross-validation}
We first consider identifiability of parameters in the model fit using cross-validation in Section \ref{subsec:karstens_mod_spec}. For sample $i$ in the training set, the mean model is
\begin{align*} \
\boldsymbol{\mu}_i = \exp(\gamma_i)[\mathbf{p}^0 \circ \exp(\boldsymbol{\beta)}) + \exp(\tilde{\gamma})3^{d_i}\tilde{\mathbf{p}}]
\end{align*}
where $\mathbf{p}^0 = \left( \mathbf{0}^T_{240} ~\frac{1}{8}\mathbf{1}_8^T\right)$ is the (known) true relative abundance vector of the synthetic community, 
$\exp(\tilde{\gamma})$ gives (unknown) intensity of contamination in an undiluted sample from the reference, 
$\tilde{\mathbf{p}}$ is the (unknown) contaminant relative abundance vector, and $d_i$
represents the (known) number of three-fold dilutions sample $i$ has undergone. For sample $i$ in the held-out/test set, the mean model is
\begin{align}
\boldsymbol{\mu}_i = \exp(\gamma_i)[\mathbf{p}\circ \exp(\boldsymbol{\beta}) + \exp(\tilde{\gamma} + \alpha)3^{d_i}\tilde{\mathbf{p}}]
\end{align}
where $\mathbf{p}$ is the unknown relative abundance vector, 
and $\exp(\tilde{\gamma} + \alpha)$ is the (unknown) intensity of
contamination in an undiluted test set sample. 

The inclusion of a contaminant profile requires additional identifiability constraints. For example, as noted in the main text, we fix $\beta_j = 0$ for all $j$ such that $p^0_j = 0$. In addition, we require that there is a taxon, call it $j^*$, that is present in the synthetic community ($p^0_{j^*} > 0$), present in the test set sample ($p_{j^*} > 0$), and absent from the contaminant profile ($\tilde{p}_{j^*} = 0$). 
Finally, we require that the ``reference taxon'' for interpreting $\exp(\boldsymbol{\beta})$ is present in the synthetic community. We may choose any reference taxon other than $j^*$. As usual, we take $\beta_{.J} = \mathbf{0}_p$. 



We proceed via contrapositive, beginning with the training fold. Suppose
\begin{align} \label{karstens_train}
 \exp(\gamma_i)[\mathbf{p}^0 \circ \exp(\boldsymbol{\beta}) + \exp(\tilde{\gamma})3^{d_i}\tilde{\mathbf{p}}] =  \exp(\gamma'_i)[\mathbf{p}^0\circ \exp(\boldsymbol{\beta}') + \exp(\tilde{\gamma}')3^{d_i}\tilde{\mathbf{p}}'].
\end{align}
Equation \eqref{karstens_train} for taxon $j^*$ gives $\exp(\gamma_i) \exp({\beta}_{j^*}) =  \exp(\gamma'_i) \exp({\beta}'_{j^*})$, and thus 
\begin{align} \label{karstens_train_divisor}
 & \frac{\mathbf{p}^0 \circ \exp(\boldsymbol{\beta})+ \exp(\tilde{\gamma})3^{d_i}\tilde{\mathbf{p}}}{\exp(\beta_{j^{\star}})} =  \frac{\mathbf{p}^0\circ \exp(\boldsymbol{\beta}') + \exp(\tilde{\gamma}')3^{d_i}\tilde{\mathbf{p}}'}{\exp(\beta_{j^{\star}}')}. 
 \end{align}
 Now, consider distinct samples $i, i'$ in the training fold such that $d_{i'} \neq d_i$. Subtracting \eqref{karstens_train_divisor} for sample $i$ from  \eqref{karstens_train_divisor} for sample $i'$, with the latter rescaled by $3^{d_{i} - d_{i'}}$, and cancelling $\tilde{\mathbf{p}}$ terms, yields
\begin{align}
& (1 - 3^{d_{i'} - d_i}) \frac{\mathbf{p}^0 \circ \exp(\boldsymbol{\beta})}{\exp(\beta_{j^{\star}})} = (1 - 3^{d_{i'} - d_i}) \frac{\mathbf{p}^0 \circ \exp(\boldsymbol{\beta}')}{\exp(\beta'_{j^{\star}})} \label{karstens_train_diff}
\end{align} 
and thus $\beta_j - \beta_{j^{\star}} = \beta'_{j} - \beta'_{j^{\star}} ~ \forall j$ such that $p^0_j>0$. 
Combining this with $\beta_{J} = \beta'_{J} = 0$, noting that $p^0_J >0$, and recalling that all other elements of $\boldsymbol{\beta}$ are constrained to equal zero, we have that
that $\boldsymbol{\beta} = \boldsymbol{\beta}'$. 
Plugging this into equation (\ref{karstens_train_divisor}) gives $\exp(\tilde{\gamma})3^{d_i}\tilde{\mathbf{p}} =  \exp(\tilde{\gamma}')3^{d_i}\tilde{\mathbf{p}}'$ and summing over $j$ and simplifying gives both $\tilde{\gamma}_i = \tilde{\gamma}'_i$ and $\tilde{\mathbf{p}} = \tilde{\mathbf{p}}'$.
  Since all other parameters in (\ref{karstens_train}) are now identified, we have $\gamma_i = \gamma'_i$ for $i$ in the training fold. Thus all parameters are identified for the training fold.


Turning now to the held-out test fold, we again proceed by contrapositive. $\tilde{\mathbf{p}}$, $\boldsymbol{\beta}$, and $\tilde{\gamma}$ are identifiable from the training fold and thus we consider
\begin{align} \label{karstens_cv_test}
 \exp(\gamma_i)[\mathbf{p}\circ \exp(\boldsymbol{\beta}) + \exp(\tilde{\gamma} + \alpha)3^{d_i}\tilde{\mathbf{p}}] =  \exp(\gamma'_i)[\mathbf{p}' \circ \exp(\boldsymbol{\beta})+ \exp(\tilde{\gamma} + \alpha ')3^{d_i}\tilde{\mathbf{p}}]
\end{align}
As above, we divide by the mean for taxon $j^{\star}$ on both sides to obtain
\begin{align} \label{karstens_test_divisor}
 & \frac{\mathbf{p}\circ \exp(\boldsymbol{\beta}) + \exp(\tilde{\gamma} + \alpha)3^{d_i}\tilde{\mathbf{p}}}{p_{j^{\star}} \exp(\beta_{j^{\star}})} =  \frac{\mathbf{p}' \circ \exp(\boldsymbol{\beta})+ \exp(\tilde{\gamma} + \alpha')3^{d_i}\tilde{\mathbf{p}}}{p_{j^{\star}}'\exp(\beta_{j^{\star}})}. 
 \end{align}
We now subtract \eqref{karstens_test_divisor} for sample $i$ from \eqref{karstens_test_divisor} for sample $i'$ ($d_i \neq d_{i'}$; both $i$ and $i'$ in the test fold) then simplifies to give 
 \begin{align}
\label{karstens_a_ratio}
\frac{\exp(\alpha)}{p_{j^{\star}}} = \frac{\exp(\alpha')}{p'_{j^{\star}}} 
\end{align}
Recalling that \eqref{karstens_cv_test} for taxon $j^*$ gives $\exp(\gamma_i) p_{j^{\star}} \exp(\beta_{j^{\star}})= \exp(\gamma_i')p'_{j^{\star}}\exp(\beta_{j^{\star}})$, and thus
\begin{align} \label{karstens_p_ratio}
\frac{p_{j^{\star}}}{p'_{j^{\star}}} = \frac{\exp(\gamma_i')}{\exp(\gamma_i)}.
\end{align}
Additionally, by the same argument that yielded equation (\ref{karstens_train_diff}),
\begin{align}
& (1 - 3^{d_{i'} - d_i}) \frac{\mathbf{p} \circ \exp(\boldsymbol{\beta})}{p_{j^{\star}} \exp(\beta_{j^{\star}})} =  (1 - 3^{d_{i'} - d_i}) \frac{\mathbf{p}' \circ \exp(\boldsymbol{\beta})}{p'_{j^{\star}} \exp(\beta_{j^{\star}})}, 
\end{align}
and therefore $\mathbf{p} \propto \mathbf{p}'$. Since both must sum to 1, $\mathbf{p} = \mathbf{p}'.$ Combining this with \eqref{karstens_a_ratio} gives $\alpha = \alpha'$ and \eqref{karstens_p_ratio} gives  $\gamma_i = \gamma'_i$ for $i$ in the test fold.
%
Hence the parameters in the test fold are also identified.

\subsubsection{Single source; no cross-validation}

To investigate the empirical coverage of confidence intervals for the elements of $\bold{p}$, we also considered a model where all samples are derived from a single source. No detection efficiencies are estimated in this model, which has mean 
\begin{align}
\boldsymbol{\mu}_i = \exp(\gamma_i)(\bold{p} + \exp(\tilde{\gamma}) 3^{d_i} \tilde{\mathbf{p}}).
\end{align}
We demonstrate identifiability assuming again that for known $j^{\star}$, $\tilde{p}_{j^{\star}} = 0$ and $p_{j^{\star}} >0$. Taking the same approach as above, we have that 
\begin{align} \label{referenceless_mean}
\exp(\gamma_i)(\mathbf{p} + \exp(\tilde{\gamma}) 3^{d_i} \tilde{\mathbf{p}}) = \exp(\gamma'_i)(\mathbf{p}' + \exp(\tilde{\gamma}') 3^{d_i} \tilde{\mathbf{p}}').
\end{align}
Dividing this expression by the mean for taxon $j^{\star}$, and subtracting the resulting expression for sample $i$ by the resulting expression for sample $i'$, we have
\begin{align}
\label{karstens_refless_div_simp}
(3^{d_i} - 3^{d_{i'}}) \frac{\exp(\tilde{\gamma}) \tilde{\mathbf{p}}}{p_{j^{\star}}} = (3^{d_i} - 3^{d_{i'}}) \frac{\exp(\tilde{\gamma}')\tilde{\mathbf{p}}'}{p'_{j^{\star}}}.
\end{align}
Summing over $j$ and rearranging, we obtain $\frac{\exp(\tilde{\gamma})}{\exp(\tilde{\gamma}')} = \frac{p_{j^{\star}}}{p'_{j^{\star}}}$.
Also, since $\tilde{p}_{j^{\star}} = 0$, we have $\exp(\gamma_i) p_{j^{\star}} = \exp(\gamma'_i)p'_{j^{\star}}$, and so $\exp(\gamma'_i - \gamma_i) = \exp(\tilde{\gamma} - \tilde{\gamma}')$. 
We substitute this into \eqref{referenceless_mean} to obtain
\begin{align}
&\frac{(\mathbf{p} + \exp(\tilde{\gamma}) 3^{d_i} \tilde{\mathbf{p}})}{\exp(\tilde{\gamma})} =  \frac{(\mathbf{p}' + \exp(\tilde{\gamma}') 3^{d_i} \tilde{\mathbf{p}}')}{\exp(\tilde{\gamma}')} 
\end{align}
Summing over $j$ gives $\frac{1 + \exp(\tilde{\gamma}) 3^{d_i}}{\exp(\tilde{\gamma})}  = \frac{1 + \exp(\tilde{\gamma}') 3^{d_i}}{\exp(\tilde{\gamma}')}$ and therefore $\tilde{\gamma} = \tilde{\gamma}',$ and therefore also $\gamma_i = \gamma'_i$ and $p_{j^{\star}} = p'_{j^{\star}}$. Equation \eqref{karstens_refless_div_simp} then gives $\tilde{\mathbf{p}} = \tilde{\mathbf{p}}'$ and finally $\mathbf{p} = \mathbf{p}'$ follows from \eqref{referenceless_mean}. 



\subsection{Identifiability: Estimating detectability, composition and contamination via samples of known composition} \label{subsec:identifiability_brooks}

In Section 7.1, we used data from \citet{brooks2015truth} to study the prediction error of our model as a function of the number of samples of known composition, randomly selecting sets of samples 
to treat as known. In this model, we have unknown detection efficiencies, an unknown contamination profile, an unknown contamination intensity, unknown sample detection effects, and unknown composition for some samples. We do, however, have known composition for some samples. Specifically, for sample $i$ of known composition, we have
\begin{align} \label{brooks_known}
\boldsymbol{\mu}_i = \exp(\gamma_i)\Big[ \mathbf{p}_i^0 \circ \exp(\boldsymbol{\beta}) + \exp(\tilde{\gamma})\tilde{\mathbf{p}}\Big]
\end{align}
for $\{{\gamma}_i\}$, $\boldsymbol{\beta}$, $\tilde{\gamma}$ and $\tilde{\mathbf{p}}$ unknown and $\mathbf{p}_i^0$ known. For sample $i$ of unknown composition, we have
\begin{align*}
  \boldsymbol{\mu}_i = \exp(\gamma_i)\Big[ \mathbf{p}_i \circ \exp(\boldsymbol{\beta}) + \exp(\tilde{\gamma})\tilde{\mathbf{p}}\Big]
\end{align*}
with $\{{\gamma}_i\}$ and $\mathbf{p}_i$ unknown. Critically, $\boldsymbol{\beta}$, $\tilde{\gamma}$ and $\tilde{\mathbf{p}}$ are shared across the samples of known and unknown composition. 

Consider a graph with $J$ nodes, each representing a taxon, and connect all nodes $(j, j')$ if there exists a sample $i$ of known composition such that both $p_{ij} >0$ and $p_{ij'} >0$. For identifiability in $\boldsymbol{\beta}$, we require that this graph is connected.  
For simultaneous identifiability of detection efficiencies and contamination profiles, 
we also require that all taxa are absent from at least two samples of known composition, and we assume that contamination profiles $\tilde{\mathbf{p}}$ lie in the 
interior of the $J-1$-simplex.

\subsubsection{Identifiability in samples of known composition}

We begin by demonstrating identifiability of the parameters present in the samples of known composition. Proceeding via contrapositive, 
if the mean model is not identifiable in $(\gamma_i, \boldsymbol{\beta}, \tilde{\gamma}, \tilde{\mathbf{p}})$ we must have that
\begin{align*}
\exp(\gamma_i)\Big[ \mathbf{p}_i^0 \circ \exp(\boldsymbol{\beta}) + \exp(\tilde{\gamma})\tilde{\mathbf{p}}\Big] & = 
\exp(\gamma_i')\Big[ \mathbf{p}_i^0 \circ \exp(\boldsymbol{\beta}')+ \exp(\tilde{\gamma}')\tilde{\mathbf{p}}'\Big]
\end{align*}
with $(\gamma_i, \boldsymbol{\beta},\tilde{\gamma},\tilde{\mathbf{p}})\neq (\gamma'_i, \boldsymbol{\beta}',\tilde{\gamma}',\tilde{\mathbf{p}}')$. 

We first demonstrate identifiability in $\tilde{\mathbf{p}}$. Define $\mathbf{j}(i) = \{j: p_{ij}^0 = 0\}$. Then, $\exp(\gamma_i)\exp(\tilde{\gamma})\tilde{p}_{\mathbf{j}(i)}$ $= \exp(\gamma_i')\exp(\tilde{\gamma}')\tilde{p}'_{\mathbf{j}(i)}$ and $\exp(\gamma_{i'})\exp(\tilde{\gamma})\tilde{p}_{\mathbf{j}({i'})} = \exp(\gamma_{i'}')\exp(\tilde{\gamma}')\tilde{p}'_{\mathbf{j}({i'})}$ for two samples $i$ and $i'$. 
Then, for $j \in \mathbf{j}(i) \cap \mathbf{j}(i')$, rearranging gives $\tilde{p}_j = a_i\tilde{p}'_{j} =  a_{i'}\tilde{p}'_{j}$, so if $\tilde{p}_j >0$, $a_i = a_{i'}$. 
Since we assumed that all taxa are absent from at least two samples of known composition, the desired samples $i$ and $i'$ must exist, and therefore $\tilde{\mathbf{p}} = a_i\tilde{\mathbf{p}}'$, which implies that 
$\tilde{\mathbf{p}} = \tilde{\mathbf{p}}'$ (both sides must sum to 1). Therefore, 
$\tilde{\mathbf{p}}$ is identifiable in this model. 

Before showing identifiability in the remaining parameters $(\gamma_i, \boldsymbol{\beta},\tilde{\gamma})$, we
first prove the following lemma guaranteeing that identifiability of these model parameters does not depend on which $j^{\star}$ we choose in 
defining our identifiability constraint $\beta_{j^{\star}} = 0$ on $\boldsymbol{\beta}$.

\begin{lemma} 
For any $j^{\star}, j^{\dagger} \in \{1, \dots, J\}$, given $\mathbf{p}$ and $\tilde{\mathbf{p}}$, the model specified in (\ref{brooks_known}) is identifiable in $(\boldsymbol{\gamma},\boldsymbol{\beta}, \tilde{\gamma})$ 
under the constraint $\beta_{j^{\dagger}} = 0$ if and only if it is identifiable in these parameters under constraint $\beta_{j^{\star}} = 0$.
\end{lemma}

\begin{proof}
Let $\boldsymbol{\beta}_{\dagger}$ denote an element of $G_{\dagger} := \{x \in \mathbb{R}^J: x_{j^{\dagger}} = 0\}$, and similarly for $\boldsymbol{\beta}_{\star}$ and $G_{\star}$. Note that 
$f(\mathbf{x}) = \mathbf{x} - x_{j^{\star}}$ defines a bijection from $G_{\dagger}$ into $G_{\star}$. Since it is sufficient to show identifiability for an arbitrary sample $i$, we proceed by constructing a bijection $g: \mathbb{R}\times G_{\dagger} \times \mathbb{R} \rightarrow \mathbb{R}\times G_{\star} \times \mathbb{R}$ such that 
$\mu_i(\gamma_i, \boldsymbol{\beta}_{\dagger}, \tilde{\gamma}) = \mu_i(g(\gamma_i, \boldsymbol{\beta}_{\dagger}, \tilde{\gamma}))$. For arbitrary $(\gamma_i, \boldsymbol{\beta}_{\dagger}, \tilde{\gamma}) \in  \mathbb{R}\times G_{\dagger} \times \mathbb{R} $,
\begin{align*}
\mu_i(\gamma_i, \boldsymbol{\beta}_{\dagger}, \tilde{\gamma}) 
&= \exp(\gamma_i)\Big[ \mathbf{p}_i^0 \circ \exp(\boldsymbol{\beta}_{\dagger} - \beta_{\dagger j^{\star}}) \exp (\beta_{\dagger j^{\star}}) + \exp(\tilde{\gamma})\tilde{\mathbf{p}}\Big] \\
&= \exp(\gamma_i + \beta_{\dagger j^{\star}} )\Big[ \mathbf{p}_i^0 \circ \exp(\boldsymbol{\beta}_{\dagger} - \beta_{\dagger j^{\star}})  + \exp(\tilde{\gamma} - \beta_{\dagger j^{\star}})\tilde{\mathbf{p}}\Big] \\
&= \mu_i(\gamma_i + \beta_{\dagger j^{\star}}, \boldsymbol{\beta}_{\dagger} - \beta_{\dagger j^{\star}}, \tilde{\gamma} - \beta_{\dagger j^{\star}})
\end{align*}
That is, $g(\gamma_i, \boldsymbol{\beta}_{\dagger}, \tilde{\gamma}) := (\gamma_i + \beta_{\dagger j^{\star}}, \boldsymbol{\beta}_{\dagger} - \beta_{\dagger j^{\star}}, \tilde{\gamma} - \beta_{\dagger j^{\star}})$ defines
a bijection from $ \mathbb{R}\times G_{\dagger} \times \mathbb{R}$ into $\mathbb{R}\times G_{\star} \times \mathbb{R}$ such that 
$\mu_i(\gamma_i, \boldsymbol{\beta}_{\dagger}, \tilde{\gamma}) = \mu_i(g(\gamma_i, \boldsymbol{\beta}_{\dagger}, \tilde{\gamma}))$. As such, if we have that 
\begin{align*}
  \mu_i(\gamma_i, \boldsymbol{\beta}_{\star}, \tilde{\gamma}) \neq \mu_i(\gamma_i', \boldsymbol{\beta}_{\star}', \tilde{\gamma}') \Rightarrow 
  (\gamma_i, \boldsymbol{\beta}_{\star}, \tilde{\gamma}) \neq (\gamma_i', \boldsymbol{\beta}_{\star}', \tilde{\gamma}')
  \end{align*}
under $\beta_{j^{\star}} = 0$ (for arbitrary $(\gamma_i, \boldsymbol{\beta}_{\star}, \tilde{\gamma}), (\gamma_i', \boldsymbol{\beta}_{\star}', \tilde{\gamma}') \in \mathbb{R}\times G_{\star} \times \mathbb{R}$), then the same must hold for arbitrary $(\gamma_i, \boldsymbol{\beta}_{\dagger}, \tilde{\gamma}), (\gamma_i', \boldsymbol{\beta}_{\dagger}', \tilde{\gamma}') \in \mathbb{R}\times G_{\dagger} \times \mathbb{R}$. This is because  
otherwise we could find $g^{-1}(\gamma_i, \boldsymbol{\beta}_{\dagger}, \tilde{\gamma}), g^{-1}(\gamma_i', \boldsymbol{\beta}_{\dagger}', \tilde{\gamma}') \in \mathbb{R}\times G_{\star} \times \mathbb{R}$ such that 
\begin{align*}
\mu_i(g^{-1}(\gamma_i, \boldsymbol{\beta}_{\dagger}, \tilde{\gamma})) = \mu_i(g^{-1}(\gamma_i', \boldsymbol{\beta}_{\dagger}', \tilde{\gamma}')) 
\text{ with } 
g^{-1}(\gamma_i, \boldsymbol{\beta}_{\dagger}, \tilde{\gamma}) \neq g^{-1}(\gamma_i', \boldsymbol{\beta}_{\dagger}', \tilde{\gamma}'),
\end{align*}
which would be a contradiction. Therefore, identifiability under $\beta_{j^{\star}} = 0$ implies identifiability under $\beta_{j^{\dagger}} = 0$. Finally, it is immediate that lack of identifiability under $\beta_{j^{\star}} = 0$ implies lack of identifiability under $\beta_{j^{\dagger}}$. If there exists
$(\gamma_i, \boldsymbol{\beta}_{\star}, \tilde{\gamma}), (\gamma_i', \boldsymbol{\beta}_{\star}', \tilde{\gamma}') \in \mathbb{R}\times G_{\star} \times \mathbb{R}$ with
$(\gamma_i, \boldsymbol{\beta}_{\star}, \tilde{\gamma}) \neq (\gamma_i', \boldsymbol{\beta}_{\star}', \tilde{\gamma}')$, and 
$\mu_i(\gamma_i, \boldsymbol{\beta}_{\star}, \tilde{\gamma}) = \mu_i(\gamma_i', \boldsymbol{\beta}_{\star}', \tilde{\gamma}')$, then $g(\gamma_i, \boldsymbol{\beta}_{\star}, \tilde{\gamma}), g(\gamma_i', \boldsymbol{\beta}_{\star}', \tilde{\gamma}')$ is a pair in $\mathbb{R}\times G_{\star} \times \mathbb{R}$ 
with the same property.
\end{proof}

We now show identifiability in $(\gamma_i, \boldsymbol{\beta},\tilde{\gamma})$. For $i$ such that $p_{ij} = 0$ for at least one $j$, consider the identifiability constraint $\beta_{j^{\star}} = 0$ for some $j^{\star}$ such that $p_{ij^{\star}} >0$ (which is WLOG by the above lemma). Then if 
$\mu_i(\gamma_i, \mathbf{p},\boldsymbol{\beta}, \tilde{\gamma},\tilde{\mathbf{p}}) = \mu_i(\gamma_i', \mathbf{p},\boldsymbol{\beta}', \tilde{\gamma}',\tilde{\mathbf{p}})$, 
we have
$\exp(\gamma_i)\exp(\tilde{\gamma})\tilde{p}_j = \exp(\gamma_i')\exp(\tilde{\gamma'})\tilde{p}_j$ and thus $\exp(\gamma_i)\exp(\tilde{\gamma}) = \exp(\gamma_i')\exp(\tilde{\gamma'})$, from which we have 
\begin{align*}
& \exp(-\tilde{\gamma})\Big[\exp(\boldsymbol{\beta})\circ \mathbf{p}_i + \exp(\tilde{\gamma})\tilde{\mathbf{p}}\Big] = 
\exp(-\tilde{\gamma}')\Big[\exp(\boldsymbol{\beta}')\circ \mathbf{p}_i + \exp(\tilde{\gamma}')\tilde{\mathbf{p}}\Big] \\
\Rightarrow & \exp(\boldsymbol{\beta} - \tilde{\gamma})\circ \mathbf{p}_i =  \exp(\boldsymbol{\beta}' - \tilde{\gamma}')\circ \mathbf{p}_i \text{    (cancelling } \tilde{\mathbf{p}} \text{ on both sides)} \\
\Rightarrow & \exp(-\tilde{\gamma})p_{ij} =  \exp(-\tilde{\gamma}')p_{ij} \\ 
\Rightarrow & \tilde{\gamma} =  \tilde{\gamma}' 
\end{align*}
since $p_{ij} >0$. Now choose another sample $i'$ with $p_{i'j^{\star}} > 0$ and at least one other nonzero entry of $\mathbf{p}_{i'}$ (by assumption such an $i$ exists). 
If in taxon $j^{\star}$ we have
\begin{align} \label{show_gam}
& \exp(\gamma_{i'}) \Big[\exp(\beta_{j^{\star}}) p_{i'j^{\star}} + \exp(\tilde{\gamma})\tilde{p}_{j^{\star}}\Big] = \exp(\gamma_{i'}')\Big[ \exp(\beta_{j^{\star}}')p_{i'j^{\star}} + \exp(\tilde{\gamma})\tilde{p}_{j^{\star}}\Big]  
\end{align}
this immediately implies $\gamma_{i'} = \gamma'_{i'}$ since $p_{i'j^{\star}} > 0$ and $\beta_{j^{\star}} = \beta_{j^{\star}}'$ (as both equal $0$ under our identifiability constraint). Now let $j'$ denote another taxon such that $p_{i'j'} > 0$. If we have
\begin{align} \label{show_bet}
\exp(\gamma_{i'})\Big[\exp(\beta_{j'})p_{i'j'} + \exp(\tilde{\gamma})\tilde{p}_{j'} \Big]  = \exp(\gamma_{i'})\Big[\exp(\beta_{j'})p_{i'j'} + \exp(\tilde{\gamma})\tilde{p}_{j'} \Big] 
\end{align}
this implies $\beta_{j'} = \beta_{j'}'$. For arbitrary $j, j'$, this argument
%
allows us to show identifiability in $\beta_{j'}$ given identifiability in $\beta_j$ conditional on the existance of an $i$ such that $p_{ij} >0$ and $p_{ij'} > 0$. Since we assume connectedness of the graph whose nodes are
taxa $1, \dots, J$ with an edge between two nodes $j, j'$ if $p_{ij} >0$ and $p_{ij'} >0$ for some known sample $i$, we are 
guaranteed to be able to find a path starting at taxon $j^{\star}$ and visiting each other taxon at least once such that 
for any adjacent $j, j'$ in our path, there exists a known sample $i$ with $p_{ij} >0$ and $p_{ij'} > 0$. 
Hence by induction with 
base case $\beta_{j^{\star}} = \beta'_{j^{\star}}$, we have $\boldsymbol{\beta} = \boldsymbol{\beta}'$ if for arbitrary $i$ we have $\mu_i(\gamma_i, \mathbf{p}, \boldsymbol{\beta}, \tilde{\gamma}, \tilde{\mathbf{p}}) = \mu_i(\gamma_i', \mathbf{p}, \boldsymbol{\beta}', \tilde{\gamma}, \tilde{\mathbf{p}})$. Finally, given identifiability of $\boldsymbol{\beta},\tilde{\gamma},$ and $\tilde{\mathbf{p}}$, identifiability of $\gamma_i$ for any sample $i$ of known composition is trivial. 
Hence the mean model is identifiable in all parameters that appear in the mean model for samples treated as known. 

\subsubsection{Identifiability in samples of unknown composition}
 
 We now turn our attention to parameters appearing only in samples treated as of unknown composition. For such a sample, a lack of identifiability now would entail
 \begin{align*}
\exp(\gamma_i)\Big[ \exp(\boldsymbol{\beta}) \circ \mathbf{p}_i + \exp(\tilde{\gamma})\tilde{\mathbf{p}}\Big] & = 
\exp(\gamma_i')\Big[ \exp(\boldsymbol{\beta}) \circ \mathbf{p}'_i + \exp(\tilde{\gamma})\tilde{\mathbf{p}}\Big]
\end{align*}
\noindent for $(\gamma_i,\mathbf{p}_i) \neq(\gamma'_i,\mathbf{p}'_i)$. This immediately gives us 
 \begin{align*}
\exp(\gamma_i)\Big[ \exp(\boldsymbol{\beta}) \circ \Big(\mathbf{p}_i + \exp(\tilde{\gamma})\exp(-\boldsymbol{\beta})\circ \tilde{\mathbf{p}}\Big)\Big] & = 
\exp(\gamma_i')\Big[  \exp(\boldsymbol{\beta}) \circ \Big(\mathbf{p}'_i + \exp(\tilde{\gamma})\exp(-\boldsymbol{\beta})\circ \tilde{\mathbf{p}}\Big)\Big]
\end{align*}
\noindent from which we have
 \begin{align*}
\exp(\gamma_i)\Big[ \mathbf{p}_i + \exp(\tilde{\gamma})\exp(-\boldsymbol{\beta})\circ \tilde{\mathbf{p}}\Big] & = 
\exp(\gamma_i')\Big[ \mathbf{p}'_i + \exp(\tilde{\gamma})\exp(-\boldsymbol{\beta})\circ \tilde{\mathbf{p}}\Big]
\end{align*}
Summing over $j$, we obtain
 \begin{align*}
\exp(\gamma_i)\Big[ 1+  \exp(\tilde{\gamma})\exp(-\boldsymbol{\beta})^T \tilde{\mathbf{p}}\Big] & = 
\exp(\gamma_i')\Big[ 1+ \exp(\tilde{\gamma})\exp(-\boldsymbol{\beta})^T\tilde{\mathbf{p}}\Big]
\end{align*}
\noindent which implies $\gamma_i = \gamma'_i$, which in turn implies that $\mathbf{p}_i = \mathbf{p}'_i$ for $i$ 
a sample treated as of unknown composition. Hence the model is identifiable in these parameters as well.

Note that while we have given sufficient conditions for a model with a single source of contamination, the same argument applies to multiple (including plate-specific) sources of contamination, since $\boldsymbol{\beta}$ does not vary across plate.

\section{SI: Additional details for reweighted estimator} \label{weighted_supp}

In Section 4, we introduced a weighted Poisson log-likelihood with weight for the likelihood contribution of $W_{ij}$ given by
\begin{align*}
\hat{w}_{ij} = \frac{\hat{\mu}_{ij} + 1}{\hat{\sigma}^2_{ij} + 1}
\end{align*}
where $\hat{\mu}_{ij}$ is the fitted mean for $W_{ij}$ given parameters
$\hat{\theta}$ estimated under a Poisson likelihood and read depth $W_{i\cdot}$. arising
from a model fit to $\mathbf{W}$ via a Poisson likelihood (without reweighting)
and $\hat{\sigma}^2_{ij}$ is a fitted value from a monotone regression
of squared residuals $(W_{ij} - \hat{\mu}_{ij})^2$ on fitted means $\hat{\mu}_{ij}$
(with $i = 1, \dots, n$ and $j = 1, \dots, J$). In other words, $\hat{\sigma}^2_{ij}$
is an estimate of $\text{var}(W_{ij}|W_{i\cdot}, \theta)$.

To motivate why this reweighting is reasonable, we consider the case in which $\mathbf{\theta}$ is in the interior of the parameter space $\Theta$.
In this setting we can express the Poisson MLE as a solution to the following score equations:
\begin{align*}
&\begin{cases} \sum_{i, j} \frac{1}{\mu_{ij}} [\frac{\partial }{\partial \theta_1}\mu_{ij}](W_{ij} - \mu_{ij})  &= 0 \\
\vdots & \\
\sum_{i, j} \frac{1}{\mu_{ij}} [\frac{\partial }{\partial \theta_L}\mu_{ij}](W_{ij} - \mu_{ij})  &= 0 \\
\end{cases}
\end{align*}
Equivalently, we can write
\begin{align*}
\sum_{i,j} \frac{1}{\mu_{ij}} \mathbf{g}_{ij}(\theta) = 0
\end{align*}
letting $\mathbf{g}_{ij} =  [\frac{\partial }{\partial \theta^T}\mu_{ij}](W_{ij} - \mu_{ij})$. Hence, we can view
this system of equations as a weighted sum of zero expectation terms $\mathbf{g}_{ij}$ with weights given by $\frac{1}{\mu_{ij}}$ ––
that is, one over a model-based estimate of $\text{Var}(W_{ij} - \mu_{ij})$.
In this setting, if the Poisson mean-variance relationship holds and the score equations have
a unique solution, we expect the estimator given by this solution to be asymptotically efficient \citep{mccullagh1983quasi},
whereas when a different mean-variance relationship holds, in general we expect to lose efficiency.
In contrast, when the Poisson mean-variance relationship does not hold,
we expect to be able to improve efficiency by reweighting the score equations
with a more flexible estimator of $\text{Var}(W_{ij} - \mu_{ij})$. To accomplish this,
we use a consistent estimator of $\theta$, the Poisson MLE $\hat{\theta}$,
to estimate $\boldsymbol{\mu}$ and $\text{Var}(W_{ij}|\mu_{ij})$.
Specifically, we estimate $\sigma^2(\hat{\mu}_{ij}) := \text{Var}(W_{ij}|W_{i\cdot}, X_i, Z_i,\tilde{Z}_i, \boldsymbol{\beta},\mathbf{p}, \tilde{\mathbf{p}}, \tilde{\boldsymbol{\gamma}})$ under the assumption
that $\sigma^2(\cdot)$ is an increasing function via a centered isotonic regression of $(W_{ij} - \hat{\mu}_{ij})^2$
on $\hat{\mu}_{ij}$. Weighting the log-likelihood contribution of $W_{ij}$, $l_{ij} := W_{ij}\text{log}(\mu_{ij})- \mu_{ij}$, by a
factor of $\frac{\hat{\mu}_{ij}}{\hat{\sigma}^2_{ij}}$ then yields reweighted score equations
\begin{align*}
\sum_{i,j} \frac{\hat{\mu}_{ij}}{\hat{\sigma}^2_{ij}} \frac{1}{\mu_{ij}} \mathbf{g}_{ij}(\theta) = \sum_{i,j} \frac{\hat{\mu}_{ij}}{\mu_{ij}} \frac{1}{\hat{\sigma}^2_{ij}} \mathbf{g}_{ij}(\theta)
\end{align*}
in which each $\mathbf{g}_{ij}$ is, up to a factor of $\frac{\hat{\mu}_{ij}}{\mu_{ij}} \stackrel{p}{\rightarrow} 1$, weighted by the inverse of a flexible
estimate of $\text{Var}(W_{ij}|\mu_{ij})$.
In practice, however, the weighting above may be unstable when
$\hat{\mu}_{ij}$ and $\hat{\sigma}^2_{ij}$ are small.
Hence, we weight instead by $\frac{\hat{\mu}_{ij} + 1}{\hat{\sigma}^2_{ij} + 1}$ to preserve behavior
of weights when the estimated mean and variance are both large (where reweighting is typically
most important) and stabilizes them when these quantities are small.

\section{SI: Supporting theory for proposed model and estimators}

Throughout this section, we will use the following notation:
\begin{itemize}
\item $\mathbf{W}_i = (W_{i1}, \dots, W_{iJ})$: a measured outcome of interest in sample $i$ across taxa $j = 1, \dots, J$. We also use $\mathbf{W}$ without subscript $i$ where this does not lead to ambiguity
\item $\mathbf{X}_i$ here denotes covariates $(Z_i,X_i, \tilde{Z}_i)$ described in the main text
\item $\mathcal{W}$: the support of $\mathbf{W} = (W_1, \dots, W_J)$
\item $\mathcal{X}$: the support of $\mathbf{X} = (X_1, \dots, X_p)$
\item $\mathcal{W}_{\lilsigma}$: the support of $W_{\lilsigma} := \sum_{j = 1}^J W_j$
\item $v$: a weighting function from $\mathcal{W}_{\lilsigma}\times \mathcal{X}$ into $\mathbb{R}_{>0}^{J}$. For simplicity of
notation, we frequently suppress dependence on $W_{i\lilsigma}$ and $\mathbf{X}_i$ and write $v_{ij}$ to indicate $v_j(W_{i\lilsigma}, \mathbf{X}_i)$
\item $\hat{v}_n$: an empirical weighting function estimated from a sample of size $n$
\item $\theta$: unknown parameters $(\mathbf{p}, \boldsymbol{\beta}, \tilde{\mathbf{p}}, \tilde{\boldsymbol{\gamma}},\boldsymbol{\alpha})$; we denote the true value with $\theta_0$
\item $\boldsymbol{\mu}_{\theta} = (\mu_{\theta 1} \dots \mu_{\theta J})$: a parametrization of the mean model given in equation (5) in main text; $\mathbb{E}[\mathbf{W}|\mathbf{X}, \gamma, \theta] = \text{exp}(\gamma)\boldsymbol{\mu}_{\theta}(\mathbf{X})$; when
unambiguous, we suppress dependence on $\mathbf{X}$ and write $\boldsymbol{\mu}_{\theta}$; we also use $\boldsymbol{\mu}_{\theta \lilsigma}$ to denote $\sum_{j = 1}^J \mu_{\theta j}$
\item $M^v_n(\theta)$: profile log-likelihood under weighting function $v$, evaluated at $\theta$ on a sample of size $n$
\item $M^v(\theta)$: expected profile log-likelihood under weighting function $v$, evaluated at $\theta$
\item $m_{\theta}^v$: the profile log-likelihood under weighting function $v$ as a function from $\mathcal{W}\times \mathcal{X}$ into $\mathbb{R}$; $M^v_{\theta} = \mathbb{E}_{\mathbf{W}, \mathbf{X}}m^v_{\theta}(\mathbf{W},\mathbf{X}) := Pm^v_{\theta}$, and similarly we can express $M^v_n(\theta)$ in terms of $m^v_{\theta}$ and the empirical measure $\mathbb{P}_n$: $M^v_n(\theta) = \mathbb{P}_n m^v_{\theta}$
\item $L^{\infty}(\mathcal{F})$: the set of all uniformly bounded real functions on $\mathcal{F}$
\end{itemize}

\subsection{Assumptions}

\begin{enumerate}[(A)]
\item We draw pairs $(\mathbf{W}, \mathbf{X}) \stackrel{\text{iid}}{\sim} P_{\theta_0}$  where
$\mathbf{W}$ has closed, bounded support $\mathcal{W} \subset \mathbb{R}_{\geq 0}^J$ and
$\mathbf{X}$ has closed, bounded support $\mathcal{X} \subset \mathbb{R}^p$.
\item Letting $\theta$ denote $(\mathbf{p},\boldsymbol{\beta}, \tilde{\mathbf{p}}, \tilde{\boldsymbol{\gamma}},\boldsymbol{\alpha})$, for a
set of known functions from $\mathcal{X}$ to $\mathbb{R}^J_{\geq 0}$ $\{\boldsymbol{\mu}_{\theta}:\theta \in \Theta\}$ we
have that
$\mathbb{E}[\mathbf{W}|\mathbf{X}, W_{\lilsigma}] = W_{\lilsigma} \frac{\boldsymbol{\mu}_{\theta_0}(\mathbf{X})}{\boldsymbol{\mu}_{\theta_0 \lilsigma}(\mathbf{X})}$
where $\theta_0 \in \Theta \subset \mathbb{R}^d$,
with $\mu_{\theta}(x)$ differentiable in $\theta$ for all $x \in \mathcal{X}$ and for each fixed $\theta \in \Theta$, $\mu_{\theta}(x)$ a bounded function on $\mathcal{X}$.
 \item For almost all $\mathbf{x} \in \mathcal{X}$, $\text{Pr}([\sum_{j = 1}^J W_j] > b | \mathbf{X} = x) = 1$ for some $b>0$.
\end{enumerate}

\noindent \textbf{Note}: while the form of the mean model given above differs somewhat from the
presentation in the main text, it in fact implies the form
in the main text if we introduce random variable $\Gamma$ and
let $\mathbb{E}[W_{\lilsigma}|\mathbf{X} = \mathbf{x}, \Gamma = \gamma] = \text{exp}(\gamma) \boldsymbol{\mu}_{\theta_0\lilsigma}$.
However, since this construction in terms of $\Gamma$ is not necessary for the results that follow,
we omit it.

%
%
%
%
%

\subsection{Form of profile log-likelihood}

We first derive the form of a log-likelihood in which nuisance parameters $\{\gamma_i\}_{i = 1}^n$, have
been profiled out. We characterize population analogue of this log-likelihood.
The form of this profile log-likelihood is as follows:
\begin{align}
M^v_n(\theta) :&= \frac{1}{n} \sum_{i = 1}^n \text{sup}_{\gamma_i \in \mathbb{R}} \Big[
\sum_{j = 1}^J v_{ij}\Big( W_{ij} \text{log}[ \text{exp}(\gamma_i) \mu_{\theta j}(X_i)] - \text{exp}(\gamma_i) \mu_{\theta j}(X_i) \Big)\Big] \\
&=  \frac{1}{n} \sum_{i = 1}^n
\sum_{j = 1}^J \Big[  v_{ij} \Big( W_{ij}\text{log}[ \frac{ \mathbf{v}_i \cdot \mathbf{W}_{i}}{ \mathbf{v}_i\cdot \boldsymbol{\mu}_{\theta}} \mu_{\theta j} ]
-  \frac{ \mathbf{v}_i \cdot \mathbf{W}_{i}}{ \mathbf{v}_i\cdot \boldsymbol{\mu}_{\theta}} \mu_{\theta j}\Big)\Big]
\end{align}
where we suppress dependence on $\mathbf{X}_i$ for simplicity in the second row. We derive the
profile likelihood in the second row via differentiation with respect to $\gamma_i$; the optimum is unique by convexity of $ay - b\text{exp}(y)$ in $y$ when $a, b>0$.
 We use $\mathbf{v}_i \cdot \mathbf{W}_{i}$ to denote
$ \sum_{j = 1}^J v_{ij} W_{ij}$ and similarly for $ \mathbf{v}_i\cdot \boldsymbol{\mu}_{\theta}$.

We now allow weights $\mathbf{v}_i = (v_{i1}, \dots, v_{iJ})$ to be given as a (bounded positive) function of $\mathbf{X}_i$ and
$\mathbf{W}_{i\lilsigma} := \sum_{j = 1}^J W_{ij}$ and examine the population analogue $M^v(\theta)$ of of the
weighted profile log-likelihood $M^v_n(\theta)$.
\begin{align}
M^v(\theta) = &\mathbb{E}_{\mathbf{W},\mathbf{X}}\Big[ \sum_{j = 1}^J v_{ij} \Big(W_{ij}\text{log}[ \frac{ \mathbf{v}_i \cdot \mathbf{W}_{i}}{ \mathbf{v}_i\cdot \boldsymbol{\mu}_{\theta}} \mu_{\theta j}  ]
-  \frac{ \mathbf{v}_i \cdot \mathbf{W}_{i}}{ \mathbf{v}_i\cdot \boldsymbol{\mu}_{\theta}} \mu_{\theta j}\Big)\Big] \\
&\mathbb{E}_{\mathbf{W}_{\lilsigma},\mathbf{X}} \mathbb{E}_{\mathbf{W}|\mathbf{W}_{\lilsigma}, \mathbf{X}} \Big[ \sum_{j = 1}^J v_{ij}\Big( W_{ij}\text{log}[ \frac{ \mathbf{v}_i \cdot \mathbf{W}_{i}}{ \mathbf{v}_i\cdot \boldsymbol{\mu}_{\theta}} \mu_{\theta j} ]
-  \frac{ \mathbf{v}_i \cdot \mathbf{W}_{i}}{ \mathbf{v}_i\cdot \boldsymbol{\mu}_{\theta}} \mu_{\theta j}\Big)\Big] \\
& = \mathbb{E}_{\mathbf{W},\mathbf{X}}\Big[ \sum_{j = 1}^J v_{ij} W_{ij} \text{log}  \mathbf{v}_i \cdot \mathbf{W}_{i}\Big] \\
& + \mathbb{E}_{\mathbf{W}_{\lilsigma},\mathbf{X}} \Big[
\sum_{j = 1}^J v_{ij} \Big( W_{i \lilsigma} \frac{\mu_{\theta_0 j}}{\mu_{\theta_0 \lilsigma}}\text{log} \frac{\mu_{\theta j}}{\mathbf{v}_i \cdot \boldsymbol{\mu}_{\theta}} -
W_{i \lilsigma} \frac{\mathbf{v}_i \cdot \boldsymbol{\mu}_{\theta_0}}{\mu_{\theta_0 \lilsigma}} \frac{\mu_{\theta j}}{\mathbf{v}_i \cdot \boldsymbol{\mu}_{\theta}} \Big)\Big]\\
& = C
 + \mathbb{E}_{\mathbf{W}_{\lilsigma},\mathbf{X}} \Big[W_{i\lilsigma} \frac{\mathbf{v}_i \cdot \boldsymbol{\mu}_{\theta_0}}{\mu_{\theta_0\lilsigma}}
\sum_{j = 1}^J v_{ij} \Big( \frac{\mu_{\theta_0 j}}{\mathbf{v}_i \cdot \boldsymbol{\mu}_{\theta_0}} \text{log} \frac{\mu_{\theta j}}{\mathbf{v}_i \cdot \boldsymbol{\mu}_{\theta}} - \frac{\mu_{\theta j}}{\mathbf{v}_i \cdot \boldsymbol{\mu}_{\theta}} \Big)\Big]
\end{align}
We note that the term in line 5 above depends on $\theta_0$ but not $\theta$; accordingly, we represent it with constant $C$ on line 7.

\subsection{Optimizer of profile likelihood}

We now show that, under a suitable identifiability condition, a weak condition on $\mathbf{W}$, and
a condition on weighting function $v$, that if the mean model given in assumption B holds at $\theta = \theta_0$, then
 the unique optimizer of population criterion $M^v(\theta)$ is $\theta_0$.

The additional conditions we need are as follows:

 \begin{enumerate}[(A)]
  \setcounter{enumi}{3}
 \item For all $\theta, \theta' \in \Theta$, we have that, for any $a \in \mathbb{R}^{+}$, $\theta \neq \theta' \Rightarrow \boldsymbol{\mu}_{\theta}(\mathbf{x}) \neq a \boldsymbol{\mu}_{\theta'}(\mathbf{x})$ holds for all $\mathbf{x} \in A \subset \mathcal{X}$ with $P_{\mathbf{X}}(A) > 0$.
  \item Weighting function $v: \mathcal{X}\times \mathcal{W}_{\lilsigma} \rightarrow \mathbb{R}_{>0}^J$,
 where $ \mathcal{W}_{\lilsigma}$ is the support of $W_{\lilsigma}$, is continuous and bounded.
 \end{enumerate}

 We will use the following simple lemma:

\begin{lemma} For every $a \geq 0$, the function defined by $f_a(b) := a\text{log}(b) - b$ is uniquely maximized at $b = a$
(defining $0\text{log}0 := 0$ and letting $a\text{log}0 = -\ ty$ for every $a>0$).
\end{lemma}

\begin{proof}
First consider the case $a >0$.
Since $f_a(0) = -\infty$ in this case and $f_a$ is finite for all $b>0$, the optimum cannot occur at $b = 0$.
Over $b \in \mathbb{R}^{+}$, $\frac{\partial^2}{\partial b^2} f = -\frac{a}{b^2} <0$, so $f_a$ is strictly convex over
$\mathbb{R}^{+}$ and hence takes a unique optimum. Setting $\frac{\partial}{\partial b} f = \frac{a}{b} - 1 = 0$ gives us that
the optimum occurs at $b = a$.

When $a = 0$, $f_a(b) = \begin{cases} 0 & \text{ if } b = 0 \\ -b & \text{ if } b > 0 \end{cases}$, so $f_a$ is optimized at $0$ since
$-b < 0$ when $b >0$.
\end{proof}

\begin{theorem}
Suppose that conditions (A) - (D) are met. Then for any weighting function satisfying (E), the criterion
$M^v(\cdot)$ defined above is uniquely optimized at $\theta_0$.
\end{theorem}

\begin{proof}
 From above we have the form of the population criterion $M^v(\theta)$:
\begin{align}
M^v(\theta) = C
 + \mathbb{E}_{\mathbf{W}_{\lilsigma},\mathbf{X}} \Big[W_{\lilsigma} \frac{\mathbf{v} \cdot \boldsymbol{\mu}_{\theta_0}}{\mu_{\theta_0\lilsigma}}
\sum_{j = 1}^J v_{ij} \Big( \frac{\mu_{\theta_0 j}}{\mathbf{v}_i \cdot \boldsymbol{\mu}_{\theta_0}} \text{log} \frac{\mu_{\theta j}}{\mathbf{v}_i \cdot \boldsymbol{\mu}_{\theta}} - \frac{\mu_{\theta j}}{\mathbf{v}_i \cdot \boldsymbol{\mu}_{\theta}} \Big)\Big]
\end{align}
For each fixed pair $(\mathbf{x},w_{\lilsigma}) \in \text{supp}(\mathbf{X}, W_{\lilsigma})$, denote by $\frac{\mu_{\theta_0 j}}{\mathbf{v} \cdot \boldsymbol{\mu}_{\theta_0}}(\mathbf{x},w_{\lilsigma})$
the function $\frac{\mu_{\theta_0 j}(\mathbf{x})}{\mathbf{v}(\mathbf{x},w_{\lilsigma}) \cdot \boldsymbol{\mu}_{\theta_0}(\mathbf{x})}$. Then
\begin{align}
h^v_j(\mathbf{x},w_{\lilsigma}, \theta; \theta_0) :=  \frac{\mu_{\theta_0 j}}{\mathbf{v} \cdot \boldsymbol{\mu}_{\theta_0}}(\mathbf{x},w_{\lilsigma}) \text{log} \frac{\mu_{\theta j}}{\mathbf{v} \cdot \boldsymbol{\mu}_{\theta}}( \mathbf{x},w_{\lilsigma})
- \frac{\mu_{\theta j}}{\mathbf{v} \cdot \boldsymbol{\mu}_{\theta}}(\mathbf{x},w_{\lilsigma})
\end{align}
is maximized when $ \frac{\mu_{\theta j}}{\mathbf{v}_i \cdot \boldsymbol{\mu}_{\theta}}(\mathbf{x},w_{\lilsigma}) = \frac{\mu_{\theta_0 j}}{\mathbf{v}_i \cdot \boldsymbol{\mu}_{\theta_0}}(\mathbf{x},w_{\lilsigma})$ by
Lemma 1.
\end{proof}

Before proceeding, we show that $-\infty < M^{v}(\theta_0) < \infty$. By definition, we have
\begin{align}
M^v(\theta_0) &= \mathbb{E}_{\mathbf{W},\mathbf{X}} \text{sup}_{\gamma \in \mathbb{R}} \sum_{j = 1}^J v_j\big(W_j \text{log}\big[ \text{exp}(\gamma) \frac{\mathbf{v}\cdot \mathbf{W}}{\mathbf{v}\cdot\boldsymbol{\mu}_{\theta_0}} \mu_{\theta_0 j} \big] - \text{exp}(\gamma) \frac{\mathbf{v}\cdot \mathbf{W}}{\mathbf{v}\cdot\boldsymbol{\mu}_{\theta_0}} \mu_{\theta_0 j}\big) \\
& \geq
\mathbb{E}_{\mathbf{W},\mathbf{X}} \sum_{j = 1}^J v_j\big(W_j \text{log}\big[ \frac{\mathbf{v}\cdot \mathbf{W}}{\mathbf{v}\cdot\boldsymbol{\mu}_{\theta_0}} \mu_{\theta_0 j} \big] - \frac{\mathbf{v}\cdot \mathbf{W}}{\mathbf{v}\cdot\boldsymbol{\mu}_{\theta_0}} \mu_{\theta_0 j}\big) \text{ (setting }\gamma = 0)\\
\label{garbage term} & = \mathbb{E}_{\mathbf{W},\mathbf{X}} \sum_{j = 1}^J v_j W_j \text{log} \mathbf{v}\cdot\mathbf{W}  \\
& + \label{notgarbage}
\mathbb{E}_{W_{\lilsigma}, \mathbf{X}} \frac{\mathbf{v} \cdot \boldsymbol{\mu}_{\theta_0}}{\mu_{\theta_0 \lilsigma}} W_{\lilsigma}
\sum_{j = 1}^J\big[ \frac{\mu_{\theta_0 j}}{\mathbf{v}\cdot \boldsymbol{\mu}_{\theta_0}} \text{log} \frac{\mu_{\theta_0 j}}{\mathbf{v}\cdot \boldsymbol{\mu}_{\theta_0}} - \frac{\mu_{\theta_0 j}}{\mathbf{v}\cdot \boldsymbol{\mu}_{\theta_0}} \big]
\end{align}
The term in line (\ref{garbage term}) is equal to $\mathbb{E}_{\mathbf{W},\mathbf{X}} \mathbf{v}\cdot\mathbf{W}\text{log} \mathbf{v}\cdot\mathbf{W}$, which as the integral of a bounded function over a bounded domain is finite.
By assumption (C), we must have $\sum_{j} \mu_{\theta_0 j}(\mathbf{X}) >0$ almost surely, so the term $\frac{\mathbf{v} \cdot \boldsymbol{\mu}_{\theta_0}}{\mu_{\theta_0 \lilsigma}} W_{\lilsigma}$ in line (\ref{notgarbage}) is almost surely bounded by boundedness of $\mu_{\theta}$, $v$, and $\mathbf{W}$. Inside
the sum in this line, we have terms of the form $a\text{log}a - a$, which is a bounded function on any bounded set in $\mathbb{R}_{\geq 0}$. Hence line (\ref{notgarbage}) is an integral of a bounded function over a bounded domain
and so is also finite, so $M^v(\theta_0) > -\infty$.

Similarly,
\begin{align}
M^v(\theta_0) &= \mathbb{E}_{\mathbf{W},\mathbf{X}} \text{sup}_{\gamma \in \mathbb{R}} \sum_{j = 1}^J v_j\big(W_j \text{log}\big[ \text{exp}(\gamma) \frac{\mathbf{v}\cdot \mathbf{W}}{\mathbf{v}\cdot\boldsymbol{\mu}_{\theta_0}} \mu_{\theta_0 j} \big] - \text{exp}(\gamma) \frac{\mathbf{v}\cdot \mathbf{W}}{\mathbf{v}\cdot\boldsymbol{\mu}_{\theta_0}} \mu_{\theta_0 j}\big) \\
&\leq \mathbb{E}_{\mathbf{W},\mathbf{X}} \sum_{j = 1}^J v_j\big(W_j \text{log}W_j - W_j) \text{ by Lemma 1} \\
& < \infty \text{ since } \sum_{j  = 1}^J v_j\big(W_j \text{log}W_j - W_j) \text{ is bounded on } \mathcal{W}\times \mathcal{X}\times\mathcal{G}
\end{align}
Hence $ -\infty < M^v(\theta_0) < \infty$, which guarantees that the difference in the following
argument is not of the form $\infty - \infty$.

Now, for any $\theta \in \Theta$ with $\theta \neq \theta_0$, we have
\begin{align*}
&M^v(\theta) - M^v(\theta_0) \\
 =& \int_{ A} \int w \frac{\mathbf{v} \cdot \boldsymbol{\mu}_{\theta_0}}{\mu_{\theta_0\lilsigma}}  \sum_{j = 1}^J \Big[ h^v_j(\mathbf{x},w_{\lilsigma}, \theta; \theta_0)   - h^v_j(\mathbf{x},w_{\lilsigma}, \theta_0; \theta_0) \Big] dP_{W_{\lilsigma} |\mathbf{X}}(w_{\lilsigma}) dP_{\mathbf{X}}(\mathbf{x}) \\
+ & \int_{A^C} \int w\frac{\mathbf{v} \cdot \boldsymbol{\mu}_{\theta_0}}{\mu_{\theta_0\lilsigma}}  \sum_{j = 1}^J \Big[ h^v_j(\mathbf{x}, w_{\lilsigma},\theta; \theta_0)  - h^v_j(\mathbf{x},w_{\lilsigma}, \theta_0; \theta_0)\Big]
 dP_{W_{\lilsigma} |\mathbf{X}}(w_{\lilsigma}) dP_{\mathbf{X}}(\mathbf{x})  \\
 \leq & \int_{A}  \int w\frac{\mathbf{v} \cdot \boldsymbol{\mu}_{\theta_0}}{\mu_{\theta_0\lilsigma}}  \sum_{j = 1}^J \Big[ h^v_j(\mathbf{x},w_{\lilsigma}, \theta; \theta_0)  - h^v_j(\mathbf{x},w_{\lilsigma}, \theta_0; \theta_0) \Big]
dP_{W_{\lilsigma} |\mathbf{X}}(w_{\lilsigma}) dP_{\mathbf{X}}(\mathbf{x}) ~~~ (\boldsymbol{\star})\\
 < &~ 0
 \end{align*}
 The first inequality is a result of $\theta_0$ maximizing (not necessarily uniquely) $h^v_j$; i.e., $h^v_j(\mathbf{x}, \theta; \theta_0)  - h^v_j(\mathbf{x}, \theta_0; \theta_0) \leq 0$.
Strict inequality holds in the last line because the integrand in $(\boldsymbol{\star})$ is strictly negative, since
$h^v_j(\mathbf{x}, w_{\lilsigma},\theta; \theta_0)  - h^v_j(\mathbf{x}, w_{\lilsigma},\theta_0; \theta_0) < 0$ for $\theta \neq \theta_0$ on $A$,
and the term $w\frac{\mathbf{v} \cdot \boldsymbol{\mu}_{\theta_0}}{\mu_{\theta_0\lilsigma}}$ is a.s. strictly positive by
assumption (C) and positivity of $\mathbf{v}$.
Hence $M^v(\theta_0) > M^v(\theta)$ for all $\theta \neq \theta_0$ in $\Theta$, so $\theta_0$ is the unique maximizer of the
population criterion $M^v(\cdot)$.

\subsection{Consistency of M-estimators}

We apply theorem 5.14 of van der Vaart (1998) to show consistency of maximizers $\hat{\theta}_n^v$
of $M^v_n$ for $\theta_0$. We also show consistency of estimators
$\hat{\theta}_n^{\hat{v}_n}$, where $\{\hat{v}_n\}$ is a sequence of random continuous positive bounded
weighting functions converging uniformly in probability to $v$.

We first require the following assumption on $v$,$\hat{v}_n$, and $\boldsymbol{\mu}$:
 \begin{enumerate}[(A)]
  \setcounter{enumi}{5}
 \item  $\text{sup}_{t \in \mathcal{X}\times \mathcal{W}_{\lilsigma}} |v(t) - \hat{v}_n(t)| \stackrel{p}{\rightarrow} 0$ and every $\hat{v}_n$ is continuous, positive, and bounded.
\item There exist $\delta >0$ and $\epsilon >0$ such that $\text{min}_j ~\inf_{\mathbf{x} \in \mathcal{X}: \mu_{\theta_0 j}(x) >0} ~\frac{\mu_{\theta j}(x)}{\mu_{\theta \lilsigma}(x)} \geq \epsilon$
holds for all $\theta \in H_{\delta} := K_{\delta} \cap \Theta$ where $K_{\delta} := \{\theta: d(\theta,\theta_0)\leq \delta\}$ is a closed $\delta$-neighborhood of $\theta_0$ in $\mathbb{R}^d$. Moreover, $\mu_{\theta}(w,x)$ is
Lipschitz continuous in $\theta$ on $H_{\delta}$ for each $(w,x) \in \text{supp}(W,X)$.

\end{enumerate}

\noindent \textbf{Note}: while assumption (G) can likely be loosened, we note that in practice it is not particularly restrictive.
In particular, we emphasize that it in no way precludes $\theta_0$ from lying at the boundary of the parameter space;
rather, it guarantees that nonzero means are bounded away from zero, which allows us to select
neighborhoods of $\theta_0$ in which $m_{\theta}^v$ is bounded.

\begin{theorem}
Suppose conditions (A) through (G) are satisfied. Then for $d(\theta,\theta') = \sum_{k = 1}^p |arctan(\theta_k) - arctan(\theta'_k)|$, $Pr(d(\hat{\theta}^v, \theta_0) > \epsilon) \rightarrow 0$ for all $\epsilon > 0$
and $Pr(d(\hat{\theta}^{\hat{v}_n}, \theta_0) > \epsilon) \rightarrow 0$ for all $\epsilon > 0$.
\end{theorem}
\noindent \textbf{Proof}

We first compactify our parameter space $\Theta$ to obtain $\bar{\Theta}$ by
allowing elements of unconstrained Euclidean parameters to take values in the extended reals.

A necessary condition for theorem 5.14 is that we have $\mathbb{E} \text{sup}_{\theta \in U} m_{\theta}^v < \infty$ for
\noindent $m_{\theta}^v(\mathbf{w},\mathbf{x}) := \sum_{j = 1}^J v_{j}\big( w_{j} \text{log}\big[ \frac{\mathbf{v}\cdot \mathbf{w}}{\mathbf{v}\cdot\mu_{\theta}} \mu_{\theta_j} \big] - \frac{\mathbf{v}\cdot \mathbf{w}}{\mathbf{v}\cdot\mu_{\theta}} \mu_{\theta_j}\big)$ and a sufficiently small ball $U \in \Theta$. By Lemma 1,
$\text{sup}_{\theta \in U} m^v(\mathbf{w},\mathbf{x}) \leq \sum_{j = 1}^J v_{j}\big( w_{j} \text{log}w_j - w_j\big)$, which is bounded above since $v$ is bounded.
Hence by assumption (A), $\mathbb{E} \text{sup}_{\theta \in U} m_{\theta}^v \leq P \sum_{j = 1}^J v_{j}\big( w_{j} \text{log}w_j - w_j\big) < \infty$.
We also require $M_n^v(\hat{\theta}^v) \geq M_n^v(\boldsymbol{\theta_0}) + o_P(1)$ which is trivially satisfied since $\hat{\theta}^v$ maximizes $M_n^v$.

Then letting compact set $K = \bar{\Theta}$, we can directly apply theorem 5.14 to obtain $Pr(d(\hat{\theta}^v, \theta_0) \geq \epsilon) \rightarrow 0$ for any $\epsilon >0$.

To apply theorem 5.14 to $\hat{\theta}^{\hat{v}_n}$, we only need in addition to the above that $M_n^v(\hat{\theta}^{\hat{v}_n}) \geq M_n^v(\theta_0) + o_P(1)$.

For any
fixed $\hat{v}$, we have
\begin{align}
M_n^{\hat{v}}(\hat{\theta}^{\hat{v}}_n) &\geq M_n^{\hat{v}}(\theta_0) + o_P(1) \\
& = M_n^{v}(\theta_0) + o_P(1) + (M_n^{v}(\theta_0) - M_n^{\hat{v}}(\theta_0))
\end{align}
However, the term $(M_n^{v}(\theta_0) - M_n^{\hat{v}}(\theta_0)) = o_P(1)$ if we let
$\hat{v} = \hat{v}_n$ since, letting $l_{ij}(\theta) := W_j \text{log}\frac{\mu_{\theta j}}{\mu_{\theta \lilsigma}} - \frac{\mu_{\theta j}}{\mu_{\theta \lilsigma}}$,
\begin{align}
| M_n^{\hat{v}_n}(\theta_0) - M_n^v(\theta_0)| & = \Big|\frac{1}{n} \sum_{i = 1}^n \sum_{j = 1}^J (\hat{v}_{n,ij} - v_ij)l_{ij}(\theta_0)\Big| \\
&\leq \text{sup}_{t \in \mathcal{X}\times \mathcal{W}_{\lilsigma}} |\hat{v}_n(t) - v(t)|  \frac{1}{n} \sum_{i = 1}^n \sum_{j = 1}^J |l_{ij}(\theta_0)| \stackrel{p}{\rightarrow} 0
\end{align}
since $\text{sup}_{t \in \mathcal{X}\times \mathcal{W}_{\lilsigma}} |\hat{v}_n(t) - v(t)| \stackrel{p}{\rightarrow} 0$.
Hence $M_n^v(\hat{\theta}^{\hat{v}_n}) \geq M_n^v(\theta_0) + o_P(1)$, so
$Pr(d(\hat{\theta}^{\hat{v}_n}, \theta_0) \geq \epsilon) \rightarrow 0$ for any $\epsilon >0$.

\subsection{Convergence in distribution of M-estimators}

\begin{theorem}
If assumptions A - E are met,  $\sqrt{n}(\hat{\theta}^v -\theta_0)$ converges in distribution to a tight limiting distribution in $\mathbb{R}^d$.
\end{theorem}

\begin{proof}
Without loss of generality, we consider $\mathcal{M}^v_{\delta} := \{m_{\theta}^v : \theta \in H_{\delta}\}$ for $H_{\delta}$ given in assumption G (since by theorem 2, with probability approaching $1$,
$\hat{\theta}_n^{v} \in H_{\delta}$ as $n$ increases without bound).

We begin by considering criterion $m_{\theta}$ without weighting.

Fix $(w, x) \in \text{supp}(W,X)$. We then have, for any $\theta \in H_{\delta}$,

\begin{align}
m_{\theta}(w,x) & = \sum_{j = 1}^J w_j \text{log} \frac{\mu_{\theta j}(x)}{\mu_{\theta \lilsigma}(x)} - \frac{\mu_{\theta j}(x)}{\mu_{\theta \lilsigma}(x)} 
\end{align}

By Lipschitz continuity of $\mu_{\theta}(x)$ in $\theta$ over $H_{\delta}$, $m_{\theta}(w,x)$ is a Lipschitz continuous function at all $\theta$ such that all elements of
$\mu_{\theta}(x)$ are nonzero since $g(z) := a \text{log}(z) - z$ is Lipschitz continuous for $z \geq \epsilon > 0$ and bounded $a \geq 0$. When one or more elements
of $\mu_{\theta}(w,x)$ are zero, they must be $0$ identically over all $\theta \in H_{\delta}$; otherwise by continuity of $\mu_{\theta}(w,x)$
in $\theta$, we are able to select $\theta^{\star} \in H_{\delta}$ such that $\mu_{\theta^{\star} j}(w, x)$ is positive but arbitrarily close to zero,
contradicting assumption G.

Since if $\mathbb{E}W_j = 0$, we must have $W_j = 0$ with probability 1 (on account of $W_j$ being a nonnegative random variable),
except possibly on a set with measure $0$, we have
\begin{align}
m_{\theta}(w,x) & =  \sum_{j = 1}^J \mathbf{1}_{\mu_{\theta_0 j}(x) > 0}\big[  w_j \text{log} \frac{\mu_{\theta j}(x)}{\mu_{\theta \lilsigma}(x)} \big] - \frac{\mu_{\theta j}(x)}{\mu_{\theta \lilsigma}(x)} \big]
\end{align}

Accordingly, $m_{\theta}(w,x)$ is Lipschitz continuous in $\theta$ for all $\theta \in H_{\delta}$. Hence
$m^{v}_{\theta}(w,x)$, as a weighted sum of Lipschitz continuous functions (with bounded weights), must also be Lipschitz continuous $\theta$ for all $\theta \in H_{\delta}$
as well.

We now use a bracketing argument to show that $\mathcal{M}_{\delta}$ is a Donsker class,
which we will use together with a result due to \citet{dumbgen1993nondifferentiable} to show
that $\sqrt{n}(\hat{\theta}_n^v - \theta_0)$ converges to a well-defined limiting distribution in $\mathbb{R}^d$.

By Lipschitz continuity of $m^v_{\theta}(w,x)$ in $\theta$ on $H_{\delta}$, we have
\begin{align} \label{brackets}
|m^v_{\theta}(w,x) - m^v_{\theta'}(w,x)| \leq C_{w,x} ||\theta - \theta'||
\end{align}

\noindent for some $C_{w,x} < \infty$.

Hence we can consider

\begin{align}
C(w,x)& = \inf\{C: |m^v_{\theta}(w,x) - m^v_{\theta'}(w,x)| \leq C ||\theta - \theta'|| \big| \theta, \theta' \in H_{\delta} \} \\
& = \sup_{\theta, \theta' \in H_{\delta}} f_{w,x}(\theta, \theta')
\end{align}

\noindent where
\begin{align}
  f_{w,x}(\theta, \theta') = \begin{cases}  \frac{|m^v_{\theta}(w,x) - m^v_{\theta'}(w,x)| }{||\theta - \theta'||} & \theta \neq \theta' \\
  \lim \text{sup}_{\theta \rightarrow \theta'}  \frac{|m^v_{\theta}(w,x) - m^v_{\theta'}(w,x)|}{||\theta - \theta'||} & \theta = \theta' \end{cases} \notag
\end{align}
where Lipschitz continuity of $m^v_{\theta}(w,x)$ guarantees the existence of the limit as $\theta' \rightarrow \theta$.

\noindent Hence by Lipschitz continuity of $m^v_{\theta}(w,x)$ in $\theta$ and continuity of $m^v_{\theta}(w,x)$ in $w$ and $x$,
$f_{w,x}(\theta, \theta')$ is
continuous in $w,x,\theta$, and $\theta'$, which in turn implies that $C(w,x)$ is continuous on $\text{supp}(W,X)$ and hence is
bounded on $\text{supp}(W,X)$ by compactness of $\text{supp}(W,X)$.
Since by assumption $\text{supp}(W,X)$ is bounded, this
implies that $|C(w,x)|^2$ is integrable, which is sufficient for the bracketing entropy of $M_{\delta}$ to be at most of
order $\text{log}(1/\epsilon)$ (see \citet{van2000asymptotic} example 19.7). Hence
by theorem 19.5 of \citet{van2000asymptotic}, $\mathcal{M}_{\delta}$ is
Donsker.

Accordingly, we have  $\{\sqrt{n}(\mathbb{P}_n m_{\theta}^v - P m_{\theta}^v): m_{\theta} \in M_{\delta}\}$ weakly converging
to a tight Gaussian process in $l^{\infty}(\mathcal{M}_{\delta})$ as $n \rightarrow \infty$. By proposition 1 of \citet{dumbgen1993nondifferentiable},
this, taken with Hadamard directional
differentiability of the map defined by $g(F) := \text{arg max}_{\theta \in H_{\delta}} F(\theta)$,
gives us
\begin{align}
\sqrt{n}(\hat{\theta}^v_n - \theta_0) := \sqrt{n}(g(\mathbb{P}_n m_{\theta}^v ) - g(P m_{\theta}^v))  \rightsquigarrow \mathcal{J}
\end{align}

\noindent for well-defined limiting distribution $J$ on $\mathbb{R}^d$.

%
%
\end{proof}
\begin{theorem}
$\sqrt{n}(\hat{\theta}^{\hat{v}_n} -\theta_0)$ converges in distribution to the same limit as $\sqrt{n}(\hat{\theta}^v -\theta_0)$ if
assumptions A - G are met.
\end{theorem}
\begin{proof}
\begin{align}
&\sqrt{n}(M^{\hat{v}_n}_n (\theta) - M^{\hat{v}_n}(\theta))\\ =& \sqrt{n}(M^{v}_n (\theta) - M^v(\theta) + \big[[M^{\hat{v}_n}_n (\theta) - M^{v}_n (\theta)] - [M^{\hat{v}_n}(\theta)  - M^v(\theta)]\big]) \\
=& \sqrt{n}(M^{v}_n (\theta) - M^v(\theta)) + \sqrt{n}(\mathbb{P}_n - P)m^{\hat{v}_n - v}_{\theta} \\
=&  \sqrt{n}(M^{v}_n (\theta) - M^v(\theta)) + o_{P}(1)
\end{align}
since
\begin{align}
\sqrt{n}(\mathbb{P}_n - P)m^{\hat{v}_n - v}_{\theta} & = \sqrt{n} \frac{1}{n} \sum_{i = 1}^n\sum_{j = 1}^J (\hat{v}_{n;ij} - v_{ij})l_{ij}(\theta) \\
&\leq \sqrt{n} \frac{1}{n} \sum_{i = 1}^n\sum_{j = 1}^J |l_{ij}(\theta)| ~\text{sup}_{t \in \text{supp}(\mathbf{X},W_{\lilsigma})} |\hat{v}_n(t) - v(t)| \\
& \stackrel{p}{\rightarrow} 0
\end{align}
Note that $\text{arg max}_{\theta \in \Theta} M^{\hat{v}_n}(\theta) = \text{arg max}_{\theta \in \Theta} M^{v}(\theta) = \theta_0$ by theorem 1.
\end{proof}

\section{SI: Optimization details}

\subsection{Reparametrization of barrier subproblem}

Letting $\theta^{\star}$ indicate the unknown parameters in our model after reparametrizing $\mathbf{p}$ and $\tilde{\mathbf{p}}$ as $\boldsymbol{\rho}$ and $\tilde{\boldsymbol{\rho}}$, we now have the following unconstrained
minimization problem:
\begin{align} \label{repar_penalty}
\text{arg min}_{\theta^{\star}} f_n(\theta^{\star})
+
\frac{1}{t^{(r)}} \Big[
&\sum_{k} \left( \sum_{j = 1}^{J - 1} -\rho_{kj} + J \text{log}\left[1 + \sum_{j = 1}^{J - 1} \text{exp}(\rho_{kj})\right] \right)\\
&\sum_{\tilde{k}}  \left( \sum_{j = 1}^{J - 1} -\tilde{\rho}_{\tilde{k}j} + J \text{log}\left[1 + \sum_{j = 1}^{J - 1} \text{exp}(\tilde{\rho}_{\tilde{k}j})\right] \right)
\Big]
\end{align}

\subsection{Barrier algorithm}

\vspace{1cm}
\noindent\fbox{
\noindent\parbox{\textwidth}{
\noindent\textit{Barrier Algorithm}
\begin{enumerate}
\item Initiate with value of penalty parameter $t$ set to starting value $t^{(0)}$ and values of parameters $\theta^{\star}$
equal to $\theta^{\star (0)}$. Set iteration $r = 0$.
\item Using current value  $t^{(r)}$ of $t$ and starting at parameter estimate $\theta^{\star(r)}$,
solve barrier subproblem r given in main text via Fisher scoring.
Denote the solution of this subproblem $\theta^{\star}_{(r + 1)}$ and set $t_{(r +1)} = a t_{(r)}$ for a prespecified
$a > 1$.
\item If $t_{(r + 1)} > t_{\text{max}}$ for prespecified $t_{\text{max}}$, return $\theta^{\star}_{(r + 1)}$. Otherwise set iteration $r = r + 1$ and
return to step 2.
\end{enumerate}
}}
\vspace{1 cm}

\subsection{Constrained Newton within Augmented Lagrangian Algorithm}

We calculate update steps from $\mathcal{L}_k$ given in Section 4 of the main text as follows:

\noindent \fbox{
\noindent \parbox{\textwidth}{
\noindent \textit{Constrained Newton within Augmented Lagrangian Algorithm}
\begin{enumerate}
\item Initiate with initial values $\nu^{(0)}$ and $\mu^{(0)} >0$ of penalty coefficients $\nu$ and $\mu$
\item Calculate proposed update $\mathbf{p}^{\text{update}}_{k}$ via nonnegative least squares on $\mathcal{L}_k$ using current values of $\nu$ and $\mu$
\item If $|\sum_{j}^J{p}^{\text{update}}_{kj} - 1 | < \delta$ for some prespecified tolerance $\delta > 0$, set update
direction $\mathbf{s}_k := \mathbf{p}_k - \mathbf{p}^{\text{update}}_{k}$ and proceed to step (3). Otherwise update
$\nu$ and $\mu$ via algorithm given in \citet{bazaraa2006nonlinear} (p. 496) and return to step (1).
\item Perform a line search in direction $\mathbf{s}_k$ to determine updated parameter value
$\mathbf{p}_{k}^{\text{updated}} := \mathbf{p}_k + \epsilon \mathbf{s}_k$ that decreases objective $f_n(\mathbf{p}_k)$ for some $0 < \epsilon \leq 1$.
\end{enumerate}
} }

\subsection{Quadratic approximation to $f_n(\mathbf{p}_k)$.}
In Section 4 of the main text, we specify $\mathcal{L}_k$ in terms of a quadratic approximation $Q_{k}^{(t)}$ to
objective $f_n(\mathbf{p}_k)$. In practice we construct $Q_k^{(t)}$ as a slightly modified Taylor
expansion of $f_n()$ around the current value of $\mathbf{p}_k^{(t)}$. We use the gradient of $f_n$
with respect to $\mathbf{p}_k$ in
the first order term, and in the second order term, and in place of the Hessian,
we use ($-1$ times) the Fisher information matrix in $\mathbf{p}_k$ regularized
(for numerical stability) by addition of magnitude of the gradient times an identity matrix.

\section{SI: Analysis of \citet{Costea:2017uv} data}

\subsection{Details of model specification} \label{costea_model}

\citet{Costea:2017uv} published two flow
 cytometric readings for every species in the synthetic community with the exception of \textit{V. cholerae}, for which only one reading was published.
 In all taxa save \textit{V. cholerae},
 we take the mean reading as our observation, and we include the resulting vector of readings augmented by the single reading for \textit{V. cholerae}
 as a row in $\mathbf{W}$. We anticipate that our use of mean readings represents a fairly small loss of information, as flow cytometric
 readings did not vary substantially within taxon. However, in a similar setting where multiple sets of flow cytometric
 readings across all taxa were available, we could
 include each set as a row of $\mathbf{W}$ to capture variability in these measurements.

To estimate detection effects relative to flow cytometry measurements, we specify $X_1 = \mathbf{0}$.
For $i\geq 2$, $\mathbf{X}_{i \cdot} = [1 ~\mathbf{1}_{Q} ~ \mathbf{1}_{W}]$
where $\mathbf{1}_{Q}$ is an indicator for sample $i$ being processed according to
protocol Q, and similarly for $\mathbf{1}_{W}$.

\subsection{Cross-validation design} \label{costea_cv_model}

We construct folds for our 10-fold cross-validation on \citet{Costea:2017uv} data so that,
with the exception of samples A and B, which we grouped together in a single fold,
 each fold included all observations for a given specimen. For each fold,
we fit a model in which all observations in all other folds, along with flow cytometry readings,
were treated as arising from a common specimen (as in fact they do, save for flow cytometry readings, which were
taken on specimens mixed to create the mock spike-in). We model each sample in the held-out fold
as arising from a distinct specimen of unknown composition to allow our model to estimate a different relative abundance
profile for distinct samples processed according to different protocols.

\subsection{Model summaries}
\begin{table}[H]

\caption{Point estimates and 95\% bootstrap confidence intervals for protocol-specific detection effects $\boldsymbol{\beta}$ (with reference taxon Y. pseudotuberculosis) estimated from \citet{Costea:2017uv} data} \label{Costea_beta}
\centering

\begin{tabular}{l|lll}
Taxon & Protocol H & Protocol Q & Protocol W\\
\hline

B. hansenii & -1.61 (-2.00 -- -1.16) & -1.55 (-1.75 -- -1.31) & -0.08 (-0.16 -- 0.00)\\

C. difficile & -0.18 (-0.30 -- 0.01) & -0.57 (-0.79 -- -0.41) & \phantom{-}1.23 (1.18 -- 1.28)\\

C. perfringens & \phantom{-}3.38 (3.27 -- 3.57) & \phantom{-}2.48 (2.31 -- 2.62) & \phantom{-}4.05 (4.03 -- 4.07)\\

C. saccharolyticum & -0.19 (-0.23 -- -0.16) & -0.01 (-0.12 -- 0.1) & -0.10 (-0.13 -- -0.06)\\

F. nucleatum & \phantom{-}2.37 (2.28 -- 2.44) & \phantom{-}0.14 (-0.16 -- 0.42) & \phantom{-}2.11 (2.05 -- 2.16)\\

L. plantarum & -2.62 (-2.96 -- -2.12) & \phantom{-}0.72 (0.60 -- 0.93) & \phantom{-}0.60 (0.56 -- 0.63)\\

P. melaninogenica & \phantom{-}4.17 (4.12 -- 4.2) & \phantom{-}3.88 (3.82 -- 4.04) & \phantom{-}4.25 (4.23 -- 4.27)\\

S. enterica & \phantom{-}2.49 (2.45 -- 2.51) & \phantom{-}2.74 (2.64 -- 2.79) & \phantom{-}2.48 (2.46 -- 2.51)\\

V. cholerae & \phantom{-}1.54 (1.50 -- 1.56) & \phantom{-}0.90 (0.78 -- 0.99) & \phantom{-}1.48 (1.44 -- 1.50)\\

\end{tabular}
\end{table}

Table \ref{Costea_beta} provides point estimates and marginal $95\%$ confidence intervals for the detection effects for each of
protocols H, Q, and W estimated via the full model described above. This model was fit with reference taxon
\textit{Y. pseudotuberculosis} (i.e., under the constraint that the column of $\boldsymbol{\beta}$ corresponding
to this taxon consists of $0$ entries). Hence we interpret estimates in this table in terms of degree of over- or
under-detection relative to \textit{Y. pseudotuberculosis} -- for example, we estimate that,
 repeated measurement under protocol H of samples consisting of 1:1 mixtures of \textit{B. hansenii} and
\textit{Y. pseudotuberculosis}, the mean MetaPhlAn2 estimate of
the relative abundance of \text{B. hansenii} will be $\text{exp}(-1.61) \approx 0.20$ as large
as the mean estimate of the relative abundance of \textit{Y. pseudotuberculosis}.

\section{SI: Analysis of \citet{karstens2019controlling} data} 

\subsection{Preprocessing}

We process raw read data reported by \citet{karstens2019controlling} using the DADA2 R package (version 1.20.0) \citep{callahan2016dada2}.
We infer amplicon sequence variants using the \textit{dada} function with option `pooled = TRUE' and
assign taxonomy with the \textit{assignSpecies} function using a SILVA v138 training dataset downloaded from https://benjjneb.github.io/dada2/training.html
\citep{quast2012silva}.

\subsection{Model Specification} \label{subsec:karstens_mod_spec}

We conduct a three-fold cross-validation of a model containing both contamination and detection effects.
For each held-out fold $r$, if we let $\mathbf{e}_r := (\mathbf{1}_{[\text{sample 1 in fold } r]}, \dots, \mathbf{1}_{[\text{sample n in fold } r]})^T$,
$\mathbf{d} = (3^0, \dots, 3^8)^T$, and $\circ$ indicate element-wise multiplication
then we specify the model for this fold with
\begin{align*}
\mathbf{Z} & = \begin{bmatrix} \mathbf{1} - \mathbf{e}_r & \mathbf{e}_r \end{bmatrix} \\
\tilde{\mathbf{Z}} &= \begin{bmatrix} \mathbf{d}\circ(\mathbf{1} - \mathbf{e}_r )\end{bmatrix} + \text{exp}(\tilde{\alpha})  \begin{bmatrix} \mathbf{d} \circ \mathbf{e}_r\end{bmatrix} \\
\mathbf{X} & = \vec{\mathbf{1}} \\
\tilde{\mathbf{X}} & = 0
\end{align*}
The relative abundance matrix $\mathbf{p}$ consists of two rows, the first of which is treated as fixed and known and contains the theoretical
composition of the mock community used by \citet{karstens2019controlling}. The second row is to
be estimated from observations on samples in the held-out fold. $\tilde{\mathbf{p}}$ consists of a single
row, the first 247 elements of which we treat as unknown. We fix $\tilde{p}_{248} = 0$ as an identifiability constraint --
identifiability problems arise here because all samples sequenced arise from the same specimen, we lack identifiability
over, for any choice of fixed $\tilde{\mathbf{p}}^{\star}$ and $\mathbf{p}$, the set $\{\tilde{\mathbf{p}} = a*\tilde{\mathbf{p}}^{\star} + (1 - a)\mathbf{p}: 0 \leq a \leq 1\}$.
Briefly, we do not consider the assumption $\tilde{p}_{248} = 0$ unrealistic; while in general distinguishing between
contaminant and non-contaminant taxa is challenging, it is fairly frequently the case that
choosing a single taxon \textit{unlikely} to be a contaminant is not difficult. Moreover, we anticipate
that in most applied settings, more than one specimen will be sequenced and this identifiability problem
will hence not arise.

$\boldsymbol{\beta}$ consists of a
single row, the first $J - 8 = 240$ elements of which (corresponding to contaminant taxa) are treated as fixed and known parameters equal to 0,
as we cannot estimate detection efficiencies in contaminant taxa. The following 7 elements of $\mathbf{\beta}$ are treated as fixed and
unknown (to be estimated from data), and $\boldsymbol{\beta}_{248}$ is set equal to 0 as an identifiability constraint.
$\tilde{\boldsymbol{\gamma}}$
is specified as a single unknown parameter in $\mathbb{R}$

The full model fit without detection efficiencies $\boldsymbol{\beta}$ is specified by treating $\boldsymbol{\beta}$ as
fixed and known with all elements equal to 0. We treat all samples as arising from the same specimen, so
$\mathbf{Z} = \mathbf{1}$, $\tilde{\mathbf{Z}} = \mathbf{d}$, and $\mathbf{p}$ consists of a single row treated as
an unknown relative abundance. Specifications of $\mathbf{X}$, $\tilde{\mathbf{X}}$,
$\tilde{\mathbf{p}}$, and
$\tilde{\gamma}$
are specified as above.

\subsection{Additional summaries}

\begin{table}[H]

\caption{Entries of $\hat{\boldsymbol{\beta}}$ (reference taxon L. fermentum) estimated from \citet{karstens2019controlling} data} \label{karstens_beta}
\centering
\begin{tabular}[t]{l|r}

Taxon & Estimate\\
\hline
P. aeruginosa & -1.29\\

E. coli & -0.20\\

S. enterica & -0.48\\

E. faecium & -2.09\\

S. aureus & -3.05\\

L. monocytogenes & -1.60\\

B. halotolerans & -0.74\\

\end{tabular}
\end{table}

For each taxon for which $\beta_j$ is identifiable (i.e., taxa in the mock community), our model produces a point estimate
$\hat{\beta}_j$, as shown in table \ref{karstens_beta}. (The reference taxon, \textit{L. fermentum}, for which we enforce
identifiability constraint $\beta_j = 0$, is excluded.) On the basis of this model, we
estimate that in an equal mixture of \textit{E. coli} and our reference taxon, \textit{L. fermentum} sequenced by the
method used by \citet{karstens2019controlling}, we expect on average to observe $\text{exp}(-0.20) \approx 0.82$ \textit{E. coli}
reads for each \textit{L. fermentum} read. In an equal mixture of \textit{S. aureus} and \textit{L. fermentum} similarly
sequenced, we expect on average to observe $\text{exp}(-3.05) \approx 0.047$ reads for each \textit{L. fermentum} read.

\section{SI: Simulation results based on \citet{brooks2015truth} data}

\subsection{Figures}

\begin{figure}[H]
\begin{centering}
\includegraphics[width=\textwidth]{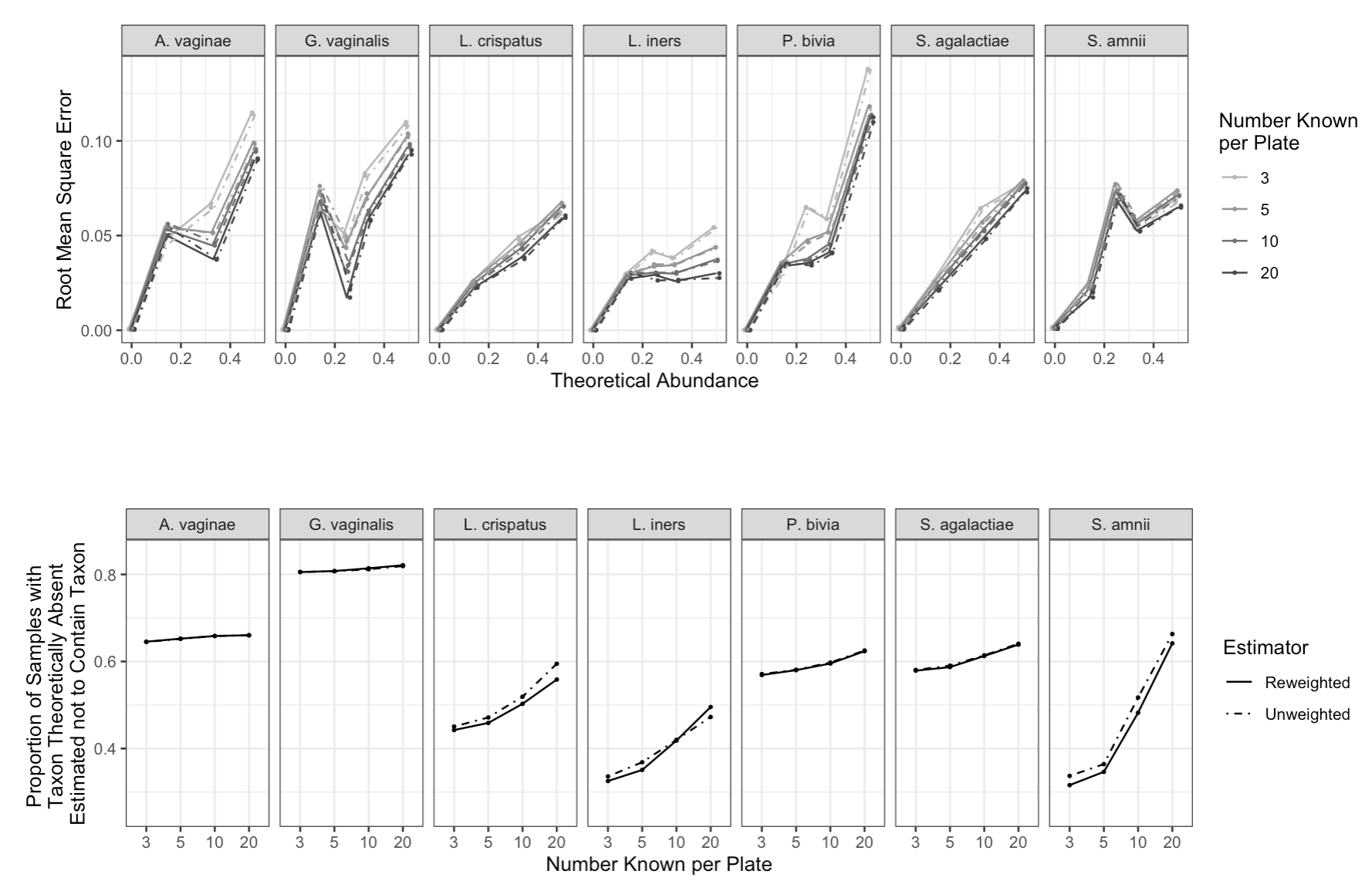}
\caption{Predictive performance of models fit on \citet{brooks2015truth} data. The
upper row includes root mean square error of relative abundance estimates by
estimator, taxon, number of samples treated as known per plate, and true relative abundance.
True relative abundance is given
on the x-axis and root mean square error is plotted on the y-axis;
for concision, true relative abundances equal to 1 are plotted at 0.
 Each column pane contains estimates for a different taxon, estimator is indicated
 with line type (solid for Poisson and dashed for weighted Poisson), and number of
 samples known per plate is indicated by color. In the lower row, proportion of
 elements of $\mathbf{P}$ truly equal to zero estimated to be equal to zero is plotted on
 the y-axis of each pane, and the x-axis gives number of samples per plate treated as known. Taxon and
 estimator are represented as in the upper row, and the proportion of nonzero elements of $\mathbf{W}$
 corresponding to zero elements of $\mathbf{P}$ for each taxon is plotted as a dotted horizontal line.} \label{brooks_figure_n}
 \label{brooks_detail}
\end{centering}
\end{figure}

Figure \ref{brooks_detail} summarizes performance of
cross-validated models fit to \citet{brooks2015truth} data.
Briefly, we observe very similar performance comparing
unweighted and weighted estimators both in terms of
root mean square error (RMSE) and proportion of elements of $\mathbf{p}$
estimated to be 0. RMSE
is generally smaller for models fit on larger training sets,
but it does not approach zero as training set size increases.
We also observe a strong relationship between RMSE and
theoretical true relative abundance, which likely reflects
a strong mean-variance relationship in the data.

We also observe generally a greater proportion of
elements of $\mathbf{p}$ estimated to be zero
with increasing training set size,
although the degree to which this occurs
depends on taxon. Weighting does not
appear to have a large impact on
this measure of predictive performance.

\section{SI: Simulations with Artificial Data}

\begin{figure}[H]
\begin{centering}
\includegraphics[width=\textwidth]{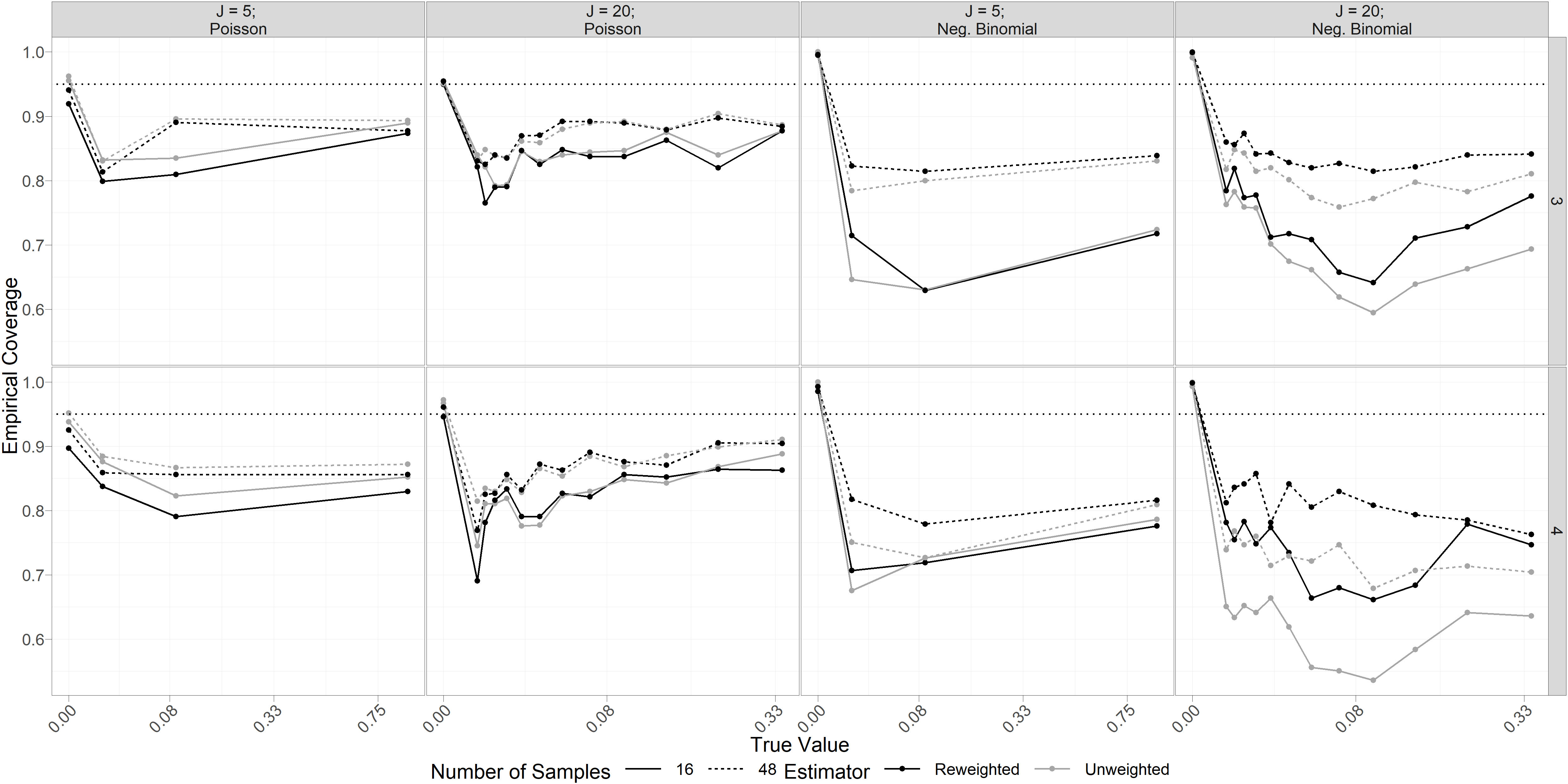}
\caption{Empirical coverage of marginal bootstrap 95\% confidence intervals
for $p_{kj}$ ($k = 3,4$) versus true value of $p_{kj}$. Coverage for
tests based on unweighted and reweighted estimators are shown in grey and black, respectively.
Sample size is indicated by line type (solid for $n = 16$ and dotted for $n = 48$).
Columns give the conditional distribution of data (Poisson or Negative Binomial) and number of taxa $J$. Rows specify which
row of $\mathbf{p}$ coverages are computed for.} \label{p_extra}
\end{centering}
\end{figure}

Figure \ref{p_extra} summarizes empirical coverage of
marginal bootstrap 95\% confidence intervals for
elements $p_{kj}$ of $\mathbf{p}$ obtained from
simulations described in the main text. As discussed
in the main text, coverage is high for $p_{kj} = 0$
but falls when $p_{kj} >0$. Unsurprisingly,
we observe higher coverage at larger sample sizes.
Coverage of intervals based on reweighted estimators
appears to be slightly lower than for unweighted estimators
when data is Poisson-distributed (lefthand columns), but
intervals using reweighted estimators substantially outperform
unweighted intervals when data is negative binomial distributed,
particularly when number of taxa $J$ is larger.

 \bibliographystyle{Chicago}

\bibliography{davids-papers-library}

\end{document}